\documentclass[12pt , titlepage]{article}

\usepackage{natbib}

\usepackage{amsmath,amssymb,amstext,amsthm,graphics, layout, lscape, longtable}
\usepackage[normalem]{ulem}
\usepackage{color}

\usepackage{url}
\usepackage{caption}
\usepackage{amssymb,amsmath,graphicx,multicol,epsfig,pictexwd,spverbatim}
\usepackage{color, colortbl}
\usepackage{multirow,multicol,dcolumn,booktabs}
\definecolor{mygray}{gray}{0.8}
\usepackage{lscape}
\usepackage[utf8]{inputenc}
\usepackage{graphicx}
\usepackage{hyperref}

\usepackage{booktabs}
\usepackage{tabularx}
\usepackage{fourier}
\usepackage{comment}
\usepackage{tablefootnote}
\usepackage{tabu}
\usepackage{lscape}
\usepackage{array}
\usepackage{caption}
\usepackage{bm}
\usepackage{xcolor,colortbl}
\usepackage{multirow}
\usepackage{makecell}
\usepackage{longtable}
\usepackage{authblk}
\usepackage{tabularray}
\usepackage{subcaption}

\usepackage{array}
\usepackage{xcolor}


\newcommand{\btheta}{\text{\boldmath{$\theta$}}}

\newcommand{\bi}{\begin{itemize}}
\newcommand{\ei}{\end{itemize}}

\newcolumntype{C}{@{\extracolsep{.4em}}c@{\extracolsep{0pt}}}%

\usepackage{setspace}
\RequirePackage[top=1in, left=1in, right=1in]{geometry}

\begin{document}

\title{Adjusting for Ascertainment Bias in Meta-Analysis of Penetrance for Cancer Risk }

\author[1]{Thanthirige Lakshika M. Ruberu}
\author[2,3]{Danielle Braun}
\author[3,2]{Giovanni Parmigiani}
\author[1*]{Swati Biswas}

\affil[1]{Department of Mathematical Sciences, University of Texas at Dallas~~~~~~~~~}
\affil[2]{Department of Biostatistics, Harvard T.H. Chan School of Public Health}
\affil[3]{Department of Data Science, Dana Farber Cancer Institute~~~~~~~~~~~~~~~~~~~~~~~~~}
\affil[*]{Correspondence to Swati Biswas, 800 W Campbell Rd, FO 35, Richardson, TX 75025. Email: swati.biswas@utdallas.edu ~~~~~~~~~~~~~~~~~~~~~~~~~}
\date{}

\maketitle

\begin{abstract}

Multi-gene panel testing allows efficient detection of pathogenic variants in cancer susceptibility genes including moderate-risk genes such as ATM and PALB2. A growing number of studies examine the risk of breast cancer (BC) conferred by pathogenic variants of such genes. A meta-analysis combining the reported risk estimates can provide an overall age-specific risk of developing BC, i.e., penetrance for a gene. However, estimates reported by case-control studies often suffer from ascertainment bias. Currently there are no methods available to adjust for such ascertainment bias in this setting. We consider a Bayesian random-effects meta-analysis method that can synthesize different types of risk measures and extend it to incorporate studies with ascertainment bias. This is achieved by introducing a bias term in the model and assigning appropriate priors. We validate the method through a simulation study and apply it to estimate BC penetrance for carriers of pathogenic variants of ATM and PALB2 genes. Our simulations show that the proposed method results in more accurate and precise penetrance estimates compared to when no adjustment is made for ascertainment bias or when such biased studies are discarded from the analysis. The estimated overall BC risk for individuals with pathogenic variants in (1) ATM is 5.77\% (3.22\%-9.67\%) by age 50 and 26.13\% (20.31\%-32.94\%) by age 80; (2) PALB2 is 12.99\% (6.48\%-22.23\%) by age 50 and 44.69\% (34.40\%-55.80\%) by age 80. The proposed method allows for meta-analyses to include studies with ascertainment bias resulting in a larger number of studies included and thereby more robust estimates.

\vspace{-0.2cm}$\\$
\end{abstract}

\section{Introduction}

Next generation DNA sequencing has expanded genetic testing for cancer-susceptibility genes from a handful of well-studied genes to a larger number (25 to 125) of genes efficiently \citep{walsh2010detection,asphaug2019cost}. The US Supreme Court's 2013 decision to overturn the patent for genetic testing has further helped in making multi-gene panel testing available at a cheaper cost \citep{Plichta2016}. Through these tests, individuals carrying pathogenic variants of cancer-susceptibility genes can be identified early on in their lives. After a carrier is identified, a key step is to determine their age-specific risk of a specific cancer, which should ideally be based on estimates from a meta-analysis. This facilitates prompt initiation of preventive measures, such as surgery or chemoprevention, to reduce the risk of associated cancers especially for those at high risk \citep{Eliade2016TheTO}. Indeed, it has been shown that panel tests have contributed to extending the life expectancy of people with hereditary cancer risk \citep{li2017multigene}.

Due to the growing usage of multi-gene panel testing, there has been a surge in the number of studies focused on estimating cancer risks conferred by cancer-susceptibility genes including moderate-risk genes such as ATM and PALB2 \citep{bonache2018multigene,coppa2018optimizing,Kurian2017,momozawa2018germline,dorling2021breast}. The reported measures of risk associated with a particular gene variant frequently vary across studies due to heterogeneity with respect to the research study design (e.g., family-based, case-control) and the type of results reported (e.g., age-specific penetrance, odds ratio (OR), relative risk (RR), standardized incidence ratio (SIR)). For clinical purposes, it is important to combine all available information on risk conferred by pathogenic variants of a specific gene from these different types of studies to determine the overall age-specific risk of a particular cancer, that is, the penetrance due to pathogenic variants of that gene. However, for this purpose, commonly employed meta-analysis techniques, fixed or random-effects models, cannot be used because they require all studies to report same type of summaries \citep{RandomMeta}.

To the best of our knowledge, \citet{Marabelli2016} developed the first meta-analysis method that has the ability to include different types of risk summaries. This is a likelihood-based fixed-effects approach wherein studies are assumed to be independent and identically distributed (iid) given the penetrance parameters shared by them. The iid assumption precludes the method to acknowledge different sources of uncertainties involved in synthesizing different types of estimates from multiple studies. Furthermore, the method was not validated using a simulation study. To overcome these limitations, we recently developed a Bayesian hierarchical random-effects model for meta-analysis to obtain penetrance estimates for a specific cancer conferred by pathogenic variants of a particular gene \citep{ruberu2023bayesian}. This approach combines results from studies that report varying types of risk measures, while also allowing for uncertainties associated with such type of synthesis. The methods of both \citet{Marabelli2016} and \citet{ruberu2023bayesian} have been applied to conduct meta-analyses to estimate breast cancer (BC) penetrance for carriers of pathogenic variants of ATM gene. Additionally, the latter has been used for estimating the age-specific BC risk for pathogenic variants of PALB2 gene for the first time in \citet{ruberupalb2}.

Studies focused on estimating cancer risk may sometimes ascertain subjects based on personal or family history of cancer, e.g., based on meeting the National Comprehensive Cancer Network (NCCN) guidelines for germline testing \citep{daly2017nccn}. If the subsequent analysis of those data does not adjust for that ascertainment condition (e.g., by conditioning the likelihood on ascertainment factors), it can lead to biased risk estimates \citep{kraft2000bias}. In particular, ascertaining individuals based on family history but not adjusting for it in the analysis may result in an overestimated risk measure due to under-representation of families with many unaffected individuals \citep{goldgar2007brca,sutton1991assessing}. 

When conducting a  meta-analysis, one possibility is to include all available studies irrespective of whether they adjusted for potential ascertainment or not, as is done in \citet{Marabelli2016}. But that can lead to biased meta-analysis estimates. Alternatively, studies that do not adjust for ascertainment can be excluded as is done in \citet{ruberu2023bayesian}. This ensures unbiased estimates but they may be less efficient compared to when all studies are included in the analysis (with some adjustment for ascertainment bias). Given that the studies subject to ascertainment adjustment do carry important information about cancer risk if they are well-designed, it is better to include them in the meta-analysis after adjustment for ascertainment bias. This is because all available information in the literature should be used for making clinical decisions about screening and prevention strategies.

From a statistical point of view, including biased studies with an appropriate bias adjustment through statistical modelling will increase the number of studies as well as the amount of information in a meta-analysis. However, this task is challenging because typically studies provide only a limited amount of summary information on the study design ascertainment. For example, \citet{hauke2018gene} conducted a case-control study to estimate the associations of heterozygous germline mutations with BC. The cases met the criteria of the German consortium for hereditary breast and ovarian cancer for germline testing, resulting in a sample enriched with family history and early-onset BC while controls were not ascertained with such criteria. The only information reported in the paper are the estimated OR, its 95\% confidence interval,  and the number of carriers among cases and controls. There is no data on the number of cases ascertained based on different criteria. In our literature review, in the context of BC risk conferred by ATM and PALB pathogenic variants, all studies that have ascertainment bias are case-control studies reporting OR. That is, we have not found any study with ascertainment bias that reported penetrance, RR, or SIR. That might be because studies reporting such measures are typically cohort studies.

A further complicating issue is a lack of studies that quantify the amount or range of bias in risk estimates for observational case-control studies with ascertainment. Nevertheless, we can make some progress by borrowing relevant information from studies based on randomized controlled trials (RCT) that focus on estimating the true intervention effect through meta-analysis while adjusting for the bias introduced by incorporating trials with high risk of bias \citep{rhodes2020adjusting,turner2009bias}. These studies are particularly useful for our purpose because they measure the true intervention effect in terms of log OR, the same scale that we use. Our goal in the current study is to extend the Bayesian meta-analysis method of \citet{ruberu2023bayesian} so that studies with reported estimates subject to ascertainment bias can be included. Our model assumes that such reported estimates over-estimate the underlying unbiased true estimate for that population and accordingly adjusts for bias in the reported estimate. We conduct a series of simulations to establish the validity of the proposed model under different settings. Finally, we apply our method to estimate age-specific risks of BC due to pathogenic variants of ATM and PALB2 genes.

\section{Methods}

A major part of the model described in this section is borrowed from \citet{ruberu2023bayesian}. We present it here for the sake of completeness as well as to ensure that our proposed extension is easy to follow.

\subsection{Likelihood Formulation}
\label{asc:likelihood}
Consider $S$ studies reporting point estimates as well as standard errors in one of four types of modalities: penetrance (age-specific risk) at certain ages, relative risk (RR), standardized incidence ratio (SIR), or odds ratio (OR). Some of these studies may have ascertainment bias. We label each study with an index $s$. Without loss of generality, we order the studies so that the first $S_1$ report penetrances, the next $S_2$ report RR, and so forth.

We assume that the cumulative penetrance $F_s(t|\kappa_s,\lambda_s)$ for study $s$ at age $t$ has the functional form given by the cumulative density function (cdf) of a Weibull distribution with shape parameter $\kappa_s$ and scale parameter $\lambda_s$ with the corresponding density function $f_s(t|\kappa_s,\lambda_s)$. Let $\btheta_s$ denote the joint parameters of study $s$ where $\btheta_s= (\kappa_s, \lambda_s)$ for unbiased studies and $\btheta_s= (\kappa_s, \lambda_s, B_s)$ for studies with ascertainment bias.

For each modality, we express the corresponding likelihood $L^{\mbox{P}},L^{\mbox{RR}},L^{\mbox{SIR}},$ and $L^{\mbox{OR}}$ of the reported results in terms of penetrance parameters as in \citet{Marabelli2016} and \citet{ruberu2023bayesian}. 
Letting $\btheta = (\btheta_1, \ldots, \btheta_S)$, we write the overall likelihood combining all the studies as

\begin{equation*}
L(\btheta)  = \prod_{s=1}^{S_1} L^{\mbox{P}} (\btheta_s) \prod_{s=S_1+1}^{\sum_{i=1}^2 S_i} L^{\mbox{RR}} (\btheta_s) \prod_{s=\sum_{i=1}^2 S_i+1}^{\sum_{i=1}^3 S_i} L^{\mbox{SIR}} (\btheta_s) \prod_{s=\sum_{i=1}^3 S_i+1}^{\sum_{i=1}^4 S_i} L^{\mbox{OR}} (\btheta_s).
\end{equation*}
We now describe the formulation of the modality-specific likelihood terms. As mentioned in the Introduction Section, all biased studies found in our literature review are case-control studies reporting OR so the adjustment for bias is described for studies reporting ORs.

\subsubsection{Age-Specific Risk (Penetrance) Estimates}
\label{asc:penet}

In this type of study, reported results include a vector $y_s^{\mbox{P}} = ( y_{1s}^{\mbox{P}}, \ldots, y_{ms}^{\mbox{P}} )$ of penetrance values at $m$ ages $(a_1, \ldots, a_m)$, along with a corresponding $m \times m$ covariance matrix $W$ quantifying estimation error. If penetrances are reported by age interval (e.g., decade) $(a_1, \ldots, a_m)$ can be the midpoints; $m$ can vary between studies. For penetrance estimation, only a subset of pathogenic variant carriers is typically used. Therefore, in our literature search we found all  studies reporting penetrance estimates adjusted for the ascertainment bias using techniques like modified segregation analysis \citep{goldgar2011rare, Antoniou2014}. Thus, our proposed method assumes that there is no ascertainment bias in studies reporting penetrance estimates.

We specify the likelihood by modeling the joint distribution of $\left\{ \mbox{logit} \left(y_{1s}^{\mbox{P}}\right), \ldots, \mbox{logit} 
\left(y_{ms}^{\mbox{P}}\right) \right\}$ given $\btheta_s$ and $W$. Ideally, the likelihood should also incorporate the sampling variability  in $W$ but we did not consider it following \citet{ruberu2023bayesian}. 
In particular, the likelihood is specified based on the following multivariate normal (MVN) distribution: 
\begin{equation*} 
\left\{ \mbox{logit} \left(y_{1s}^{\mbox{P}}\right), \ldots, \mbox{logit} 
\left(y_{ms}^{\mbox{P}}\right) \right\}
\sim \mbox{MVN} \Bigl( \bigl\{ \mbox{logit} ( F_s(a_1 | \kappa_s, \lambda_s) ),\ldots,
\mbox{logit} ( F_s(a_m | \kappa_s, \lambda_s) ) \bigr\}, W^*
\Bigr) = L^{\mbox{P}} (\btheta_s).
\end{equation*}

Here, $W^*$ can be obtained from $W$ via, say, an application of the multivariate delta method.
However, most studies report only 95$\%$ confidence intervals (CIs) of the individual components of $y_s^{\mbox{P}}$. These CIs are used to obtain the diagonal elements of $W^*$ (i.e., estimated variances). Specifically, we take the logit of the lower and upper limits of each age-specific CI, compute the width of the resulting interval, and estimate the corresponding variance by assuming normality. The covariances in the off-diagonal elements of $W^*$ are approximated using a numerical method of \citet{Marabelli2016}, which is described in details in \citet{ruberu2023bayesian}.

\subsubsection{Odds Ratio}
\label{asc:OR}
In this type of study, the reported results include a scalar 
OR estimate $y_s^{OR}$ with variance $w_s$. These studies are case-controls studies, which typically ascertain cases differently from controls. Thus if the reported OR was not obtained using a conditional analysis, it is likely to be biased. For studies with potential ascertainment bias, we assume that the reported estimate overestimates the true OR value for that particular population. That is because cases are typically selected from higher risk populations (e.g., based on family history) as compared to controls. We specify the likelihood by modeling the distribution of $\mbox{log} ( y_{s}^{\mbox{OR}} )$ given $\btheta_s$ and $w_s^*$, where $w_s^*$ is an appropriate transformation of $w_s$. We use the log transformation to facilitate the fit of the normal approximation.  For unbiased OR estimates, we posit the following distribution

\[ \log y_s^{OR} \sim N \bigg(\log (\nu_s), \; w_s^*\bigg), \] 
where $\nu_s$ is an approximation of OR in terms of penetrance. For studies with ascertainment bias, we assume 
\[ \log y_s^{OR} \sim N \bigg(\log (\nu_s) + B_s^{OR}, \; w_s^*\bigg), \] 
where $B_s^{OR}$ is the bias of study $s$ in log OR scale ($B_s^{OR}>0$). In both cases, following \citet{ruberu2023bayesian}, $\nu_s$ is expressed in terms of penetrance as
\begin{equation}\label{asc.OR}
\nu_s = \left. \frac{\int{f_{s}(a|\kappa_s, \lambda_s) q_{c1}(a)da}}{\int{f_{0}(a) q_{c0}(a)da}} \middle/\frac{\int{(1-F_{s}(a|\kappa_s, \lambda_s)) q_{h1}(a)da}}{\int{(1-F_{0}(a))q_{h0}(a)da}} \right.,
\end{equation}
where $q_{c1}$ and $q_{c0}$ are distributions of ages of onset for cases with 1 and 0 in the subscript indicating carriers and non-carriers, respectively; while $q_{h1}$ and $q_{h0}$ are the corresponding distributions of ages at inclusion in the study for healthy controls and $f_0(a)$ is the density of the penetrance of BC among non-carriers (considered to be known) with c.d.f $F_0(a)$. We assume that the age-related distributions are $q_{c1} = N(A_{c1}, V_{c1})$, $q_{c0} = N(A_{c0}, V_{c0})$, $q_{h1} = N(A_{h1}, V_{h1})$, and $q_{h0} = N(A_{h0}, V_{h0})$, whose the mean and variance parameters $A$ and $V$ are pre-specified. Note that these are unrelated to the assumed Weibull cdf at age $a$, which is used to model the probability of getting cancer by age $a$, i.e., it is conditional on age $a$.

\subsubsection{Relative Risk and Standardized Incidence Ratio} 
\label{asc:RR}
Following \citet{Marabelli2016} and \citet{ruberu2023bayesian} we model studies reporting SIR in the same manner as those reporting RR under the assumption of rare gene mutation in which case the incidence of BC in a general population is approximately equal to that in non-carriers.  

In our literature review of ATM and PALB2 pathogenic variants in BC, we did not find any RR study with ascertainment bias. However, if there is such a study, e.g., a case-control study with ascertainment bias, the bias can be modeled in the same way as for studies reporting OR. Thus, in principle, we can incorporate studies subject to ascertainment bias in addition to studies that are unbiased. Those with ascertainment bias are assumed to overestimate the underlying true RR of the study population. For both types of studies, reported results include a scalar RR estimate $y_s^{RR}$ with variance $w_{s}$. For unbiased studies, expressing the mean of the RR approximately in terms of the penetrance, we posit the following distribution:  
\begin{equation}\label{asc.RR1}
  \log y_s^{RR} \sim N \Biggl(
  \log \left ( \frac{\int{f_{s}(a|\kappa_s, \lambda_s) q_1(a)da}}{\int{f_{0}(a) q_0(a)da}}
 \right ) ,
  w_s^* \Biggr ),
\end{equation}
where $w_s^*$ is obtained similarly to the OR case and
$q_1(a) = N(A_1, V_1)$ and $q_0(a) = N(A_0, V_0)$ are distributions of age of onset among carriers and non-carriers.

For studies with ascertainment bias, we modify the mean in equation \ref{asc.RR1} as a sum of two terms --- RR in terms of the penetrance and the bias in the reported estimate and we posit the following distribution:
\begin{equation}\label{asc.RR2}
  \log y_s^{RR} \sim N \Biggl(
  \log \left ( \frac{\int{f_{s}(a|\kappa_s, \lambda_s) q_1(a)da}}{\int{f_{0}(a) q_0(a)da}}
 \right ) + B_s^{RR},
  w_s^* \Biggr ),
\end{equation}
 
where $B_s^{RR}$ is the bias of study $s$ in log RR scale. Note that $B_s^{RR}>0$.

\subsection{Prior Distributions}
\label{asc:prior}
As in \citet{ruberu2023bayesian}, we assign Gamma priors to our study-specific $\kappa_s$ and $\lambda_s$ parameters and Uniform distributions with pre-specified limits to the hyper-parameters. Specifically, we use the following hierarchical priors: $\pi(\kappa_s|a, b) =$ Gamma$(a, b), \;  \pi(\lambda_s|c, d) =$ Gamma$(c, d),\;$ where $a$ and $c$ are shape parameters, $b$ and $d$ are scale parameters, and $\pi(a|l_a, u_a) =$ U$(7.5,27.5),\;$ $\pi(b|l_b, u_b) =$ U$(0.15,0.25), \;  \pi(c|l_c, u_c) =$ U$(43,63), \;$and$ \; \pi(d|l_d, u_d) =$ U$(1.32,2.02)$. This prior set-up allows a wide spectrum of penetrance values and thus is applicable to a broad range of gene-cancer combinations. Moreover, \citet{ruberu2023bayesian} had conducted a sensitivity analysis for the choice of fixed hyper-parameters and reported that the results are robust. Additionally, we assign two different priors for $B_s^{OR}$ and $B_s^{RR}$. Specifically, $B_s^{OR} \sim$ half-normal$(\sigma=0.9)$ and $B_s^{RR} \sim$ half-normal$(\sigma=0.5)$. We choose the half-normal distribution because it is a single-parameter distribution defined on the positive real line. 

Due to the lack of quantitative information about ascertainment bias in the literature, the value of $\sigma$ parameter is derived based on bias information from RCT settings. In RCT settings, one approach uses elicited opinion on the likely extent of bias wherein experts are asked to summarize their beliefs about bias by providing a numerical range representing their uncertainty \citep{rhodes2020adjusting,turner2009bias}. We borrow information about the range of bias in log OR reported in these studies. However, as ascertainment bias in case-control studies may be more extreme and vary more across studies than the biases in RCT, we assign a half-normal prior with mean ($\sigma\sqrt{2/\pi}$) of 0.72 and SD ($\sigma\sqrt{1-(2/\pi)}$) of 0.54 for $B_s^{OR}$, which allows such possibilities (see Figure S1a). Studies reporting RR tend to report estimates that are lower and have less variability compared to studies reporting OR. Therefore, for $B_s^{RR}$ we use a prior with a lower mean of 0.40 and SD of 0.30  (see Figure S1b). Note that none of actual RR studies that we found had ascertainment bias though. Thus, allowance for bias in RR studies is just for allowing flexibility in modelling, if needed rather than for practical utility. 

\subsection{Posterior Distributions}
\label{asc:MCMC}

For estimating posterior distributions, we implement a Markov chain Monte Carlo (MCMC) algorithm. Detailed steps of the MCMC algorithm can be found in the Supplement.
We run 30,000 MCMC iterations with 15,000 burn-in. The convergence of the algorithm is assessed using Gelman-Rubin statistic and trace plots \cite[Chapters~11-12]{gelman1995bayesian}. 

Once the posterior distributions are estimated, the final meta-analysis penetrance curve is obtained using the following steps : (1) Compute $\kappa^{(t)}=a^{(t)}*b^{(t)}$ and $\lambda^{(t)}=c^{(t)}*d^{(t)}$ and (2) Use the Weibull($\kappa^{(t)},\lambda^{(t)}$) cdf at ages 40, 50, 60, 70, and 80 as penetrance estimates at the $t^{th}$ iteration. Finally, for each age, the mean of the penetrance values over all iterations are computed, which serves as the final meta-analysis estimate. Credible intervals (CrI) are also obtained using these posterior distributions of penetrances at each age. 
We carry out all analyses in statistical software system R \citep{R}.

\section{Simulation Study}
\label{asc:Sim}
\subsection{Simulation Set-up}
We conduct a simulation study to evaluate the performance of the method based on studies reporting ATM-BC associations. We have conducted a literature review and identified 30 such studies (Table \ref{asc.meta:ATM}) that we include in our actual meta-analysis to be presented in Section \ref{asc.application:ATM}. Among these studies, 20 are unbiased studies while 10 studies provide OR with ascertainment bias. Among the 20 unbiased studies, 2 report penetrance, 5 report RR/SIR, and 13 report OR. In addition to simulating under this specific combination of unbiased and biased studies reporting OR, we explore other simulation settings in which we vary the number of studies with ascertainment bias among the 23 studies reporting OR. Additionally, even though none of the actual RR studies have ascertainment bias, we also consider a setting with biased RR studies. Specifically, the following settings are considered in which there are a total of 23, 2, and 5 studies reporting OR, penetrance, and RR/SIR, respectively (except Setting 4).

\begin{itemize}
    \item Setting 1 : 10 OR studies  are subject to ascertainment bias (13 OR studies are unbiased).
    \item Setting 2 - 5 OR studies  are subject to ascertainment bias (18 OR studies are unbiased).
    \item Setting 3 - 15 OR studies  are subject to ascertainment bias (8 OR studies are unbiased).
    \item Setting 4 - 10 OR studies  are subject to ascertainment bias (13 OR studies are unbiased) and 5 additional hypothetical RR studies subject to ascertainment bias (5 unbiased RR/SIR studies remain as per Table \ref{asc.meta:ATM} and in other settings).
\end{itemize}

In all simulation settings except Setting 4, 30 studies are generated in each simulation replicate with sample sizes and type of reported risk measures (penetrance, RR, or OR) the same as those reported in the actual studies listed in Table \ref{asc.meta:ATM}. Additionally, a bias term is generated for the studies that are subject to ascertainment bias (to be further elaborated in Section ~\ref{asc:data generate}). In Setting 4, five additional studies reporting RR are generated subject to ascertainment bias making the total number of studies in this setting to be 35.

To mimic the actual studies listed in Table \ref{asc.meta:ATM}, we assume that only a subset of the studies report age-related statistics that are necessary for calculating equations ~\eqref{asc.OR}, ~\eqref{asc.RR1}, and ~\eqref{asc.RR2}. We refer to this scenario as Scenario 1. Specifically, in Settings 1 through 3, if a study listed in Table \ref{asc.meta:ATM} provides relevant age-related information, we use the simulated age-related summaries of that study. Otherwise, we assign a fixed mean of 63 and an SD of 14.00726, which are the mean and SD of the age of onset of breast cancer in the US population, as obtained from the Surveillance, Epidemiology, and End Results (SEER) program \citep{seer2}. For Setting 4, the availability of age-related summaries in the five additional hypothetical RR studies is assumed to follow the same respective pattern as in the five RR/SIR studies listed in Table \ref{asc.meta:ATM}. 

Case-control studies usually only provide the mean age at diagnosis for cases and the mean age at inclusion in the study for controls. Specific age distributions for carriers and non-carriers within cases and controls are typically not provided. Therefore, similar to \citet{ruberu2023bayesian} and \citet{Marabelli2016}, for case-control studies, we set $q_{c1}(a) = q_{c0}(a)=q_{c}(a)$ (age of onset for cases) and $q_{h1}(a) = q_{h0}(a)=g_{h}(a)$ (age at inclusion in the study for healthy controls) because these studies usually only provide the mean age at diagnosis for cases and the mean age for controls. 

To investigate the effect of age-related distributions on the final estimates, we consider an alternate simulation scenario wherein none of the studies provide pertinent age-related summaries (Scenario 2). We simulate Scenario 2 under Setting 1, which mimics the ATM-BC studies included in our meta-analysis. In this case, all age-related summaries $q_1$, $q_0$, $q_{c}$, and $g_{h}$ for all studies are assumed to have the same fixed mean and SD of $63$ and  $14.00726$, respectively. 

We compare our proposed method with two other approaches. Both use the method developed by \citet{ruberu2023bayesian} --- one approach includes studies with ascertainment bias while the other excludes them. Thus, we compare three approaches: 
 \begin{enumerate}
      \item Approach 1 - Meta-analysis includes studies with ascertainment bias. Model adjusts for the bias.
      \item Approach 2 - Meta-analysis includes studies with ascertainment bias. Model does not adjust for the bias.
      \item Approach 3 - Meta-analysis excludes studies with ascertainment bias. This results in a lesser number of studies.
  \end{enumerate}

\subsubsection{Data Generation Model}
\label{asc:data generate}
Depending on the simulation setting, we generate a total of either 30 or 35 studies in each simulation replicate. We first generate unbiased estimates for all studies and then incorporate bias into the estimates that come from studies with ascertainment bias. 
For unbiased studies, the generated estimate equals the reported estimates while for a biased study we add an extra bias term to the generated estimate to obtain the reported estimate.

Following \citet{ruberu2023bayesian}, we generate time to BC for carriers by using a Weibull$(\kappa_s,\lambda_s)$ distribution (for event) in conjunction with a censoring time sampled from $N(85,10)$. The study-specific $\kappa_s$ and $\lambda_s$ are generated from $N(4.55,0.525)$ and $N(95.25,12.375)$, respectively. Next, we describe modality-specific data generation procedure which closely follows \citet{ruberu2023bayesian} except the generation of ascertainment bias.
\\
{\bf{\emph{Age-Specific Penetrance}}}$\\$
We generate two time points for each carrier in the $s^{th}$ study: (1) Time to BC and (2) Censoring time. The minimum of these two times and age 95 (assumed to be the maximum observed age in a study), and an indicator of censoring are considered to be the observed data for each carrier. Next, we fit a Kaplan-Meier curve and get penetrance estimates and their CIs at ages 40, 50, 60, 70, and 80. Note that we do not generate non-carriers because they are not needed for estimation of penetrance. 
\\
{\bf{\emph{RR and OR}}}$\\$
First, we generate a population of 2 million consisting of carriers and non-carriers with carrier probability of 0.01. Then, BC status for each carrier is generated in the same manner as described above for studies reporting penetrance. In particular, a subject is affected if her time to BC is equal to the minimum of the three times (time to BC, censoring time, and 95) otherwise she is unaffected. Next, the process is repeated for non-carriers by generating time to BC using a truncated Weibull$(3.65, 143.2426)$ distribution, which is an approximation to BC risk estimates given by SEER. The Weibull distribution is truncated at age 185 to ensure that there are enough non-carrier cases at ages less than 80 so that the resulting RR and OR estimates are not excessively large.

For studies reporting OR, the estimate OR and its standard error are calculated for a randomly chosen sample of cases and controls with sample sizes same as those of twenty OR studies listed in Table \ref{asc.meta:ATM}. The bias terms are generated using Gamma$(2.312,3.4)$ distribution (shown in Figure S1a). This allows for a wide yet reasonable range of bias values that can be expected in actual studies (note that it may be better to exclude studies with extreme bias in a meta-analysis). In Settings 1 - 4, we generate 10, 5, 15, and 10  bias terms, respectively in each replicate. These bias terms are then added to the log of the generated OR for the studies that are assumed to have ascertainment bias and the resulting OR is the reported OR for each such study.

Finally, for studies reporting RR, the estimate RR and its standard error are obtained for a randomly chosen sample of carriers and non-carriers. The sample sizes of carriers and non-carriers are the same as those listed in Table \ref{asc.meta:ATM}. In Setting 4, we generate RR for five hypothetical studies whose sample sizes $n$  are 400, 1100, 250, 40, and 70. For these five studies, an additional bias term is sampled from Gamma$(2.04,5.83)$ distribution (shown in Figure S1b), which is then added to the log of the generated RR estimate to obtain the reported estimate. 

Note that the age at which a given subject is either diagnosed, censored, or healthy is known at this point of data generation along with their carrier status. Using the mean and variance of the corresponding ages, we construct relevant (normal) age distributions for a study that reports age-related distributions under Scenario 1. For OR studies, we set the mean and SD of ages of the controls same as the generated ones for the cases because cases and controls are often matched on age distributions. In Scenario 1, whenever a study did not report a given age-related distribution, that distribution was replaced by $N(63,14.00726)$ whereas in Scenario 2, we use $N(63,14.00726)$  for all age-related distributions for all studies.

The estimates and their standard errors obtained from all studies along with the relevant age distributions are then input to the meta-analysis model. To evaluate the results of the simulations, we need to know the true penetrance values, which are not directly available as each study has its own specific penetrance curve. So we generate a large number of $\kappa_s$ and $\lambda_s$ parameters from their respective normal distributions (which were used in data generation process as described above) to obtain a large number of Weibull($\kappa_s$, $\lambda_s$) curves. From these curves, we obtain the corresponding penetrance values at ages 40 to 80 with 10-year increment and compute their mean for each age. The mean penetrance values at these five ages serve as true values for evaluation of the simulation results. 

For each combination of simulation setting and scenario, we generate 500 replicates. After applying all three approaches to each replicate, we report average of the penetrance estimates, their root mean square error expressed as a fraction of true value of penetrance (True) i.e., RMSE/True, and coverage probabilities of $95\%$ CrIs.

\subsection{Simulation Results}

Tables ~\ref{Table1}, \ref{Table2}, \ref{Table3} and S1 show the results of applying the above-mentioned three approaches to Settings 1 - 4. The following observations stand out when we compare approaches 1 (biased studies are included and model adjusts for bias) and 2 (biased studies are included but model does not adjust for bias) at all ages and settings except age 40 under Setting 4: (1) The estimates from Approach 1 are closer to the true penetrance values compared to those from Approach 2 with the latter being over-estimate (2) Approach 2 has higher RMSE/True values (3) Approach 1 has coverage probability of $95\%$ CrIs close to 95 while Approach 2 has lower coverage at ages 70 and 80. These results underscore the importance of accounting for ascertainment bias when incorporating studies that enroll cases on the basis of personal or family history, which may increase the reported ORs.

When we compare approaches 1 and 3, we can see that (1) penetrance estimates are similar (2) RMSE/True values are lower for Approach 1 with the exception of ages 40 and 50 in Setting 4 (recall that this setting contain higher proportion of biased studies) (3) $95\%$ coverage probability at each age are close to 95 under both approaches. 
Our results suggest that incorporating studies that report biased estimates due to ascertainment bias and adjusting for bias using the model is preferable to discarding such studies from the analysis.

Table S2 contain the simulation results for Setting 1 under Scenario 2. The penetrance estimates and coverage probability of $95\%$ CrIs  at each age are close to those observed in Table \ref{Table1} (Setting 1 -- Scenario 1). At each age, RMSE/True values are lower for Setting 1 -- Scenario 2 (Table S2) compared to Setting 1 -- Scenario 1 (Table \ref{Table1}) for all approaches.

\subsection{Sensitivity Analysis}

 We investigate how sensitive the results are to the prior assumed for $B_s^{OR}$, the bias parameter used in modelling studies reporting biased OR. The prior distribution used in our method is half-normal $(\sigma=0.90)$ whose mean is 0.72. We explore two other priors: (1) half-normal $(\sigma=0.76)$ with mean of 0.61 (2) half-normal $(\sigma=1.09)$ with mean 0.87. These are chosen in such a manner that they span a reasonable range of bias in log OR scale. Figure S2 shows the densities of these different prior distributions. 
 
 Table S3 provides the results of the sensitivity analysis for Setting 1 -- Scenario 1 for Approach 1 (note that the other two approaches do not account for bias). In comparison to  Table \ref{Table1}, there is very little difference in results, establishing the robustness of the proposed approach to varying priors on $B_s^{OR}$.

\section{Applications}
\label{asc:app}

Here, we apply the proposed Bayesian method to estimate the age-specific penetrance of female BC among carriers of pathogenic variants in ATM  and PALB2 genes.

\subsection{Assumptions Specific to Analyses of ATM and PALB2 genes}

The following assumptions are made in line with \citet{ruberu2023bayesian} and \citet{Marabelli2016}. We assume $q_{c1}(a) = q_{c0}(a)=q_{c}(a)$ and $q_{h1}(a) = q_{h0}(a)=q_{h}(a)$ (these were defined after equation ~\eqref{asc.OR}). Moreover, if mean and variance of the age of healthy controls were not reported by a paper, we further assume that $q_{ci}(a) = q_{hi}(a) \text{ for } i=0,1$. Whenever a study did not report any age-related summary, we use a mean age of 63 and SD of 14.00726, the same ones that used in our simulations (obtained from SEER). For case-control studies in which no mutations were reported in controls (i.e., OR is not defined), we add $0.5$ to each cell of the $(2 \times 2)$ table to estimate OR and its SE \citep{jb1956estimation,gart1967bias}.

\subsection{Meta-Analysis of ATM-BC Penetrance}
\label{asc.application:ATM}
\subsubsection{Selection of Studies}
\label{asc.se:ATM}
We start with all 17 studies included in the meta-analysis of \citet{ruberu2023bayesian}. However, \citet{ruberu2023bayesian} included only those studies that are unbiased i.e., they excluded eligible studies that did not adjust for ascertainment criterion in their analysis. We include those excluded studies in the current meta-analysis. 

This introduces 10 additional studies reporting OR subject to ascertainment bias into our meta-analysis. Note that only OR studies had potential ascertainment bias. \citet{ruberu2023bayesian} had found the above 27 studies by Pubmed search of following keywords in the title/abstract of the articles: [‘‘ATM’’] AND [‘‘penetrance’’ OR ‘‘risk’’] AND [‘‘breast’’] up to December 20, 2021. We extend the same search to identify additional studies published up to May 10, 2023. The inclusion criteria for studies is similar to those of \citet{Marabelli2016} and \citet{ruberu2023bayesian}. In particular, we include  (1) family-based segregation analyses or epidemiological studies reporting cancer risk information, in terms of age-specific penetrance, RR, or SIR and (2) case-control studies comparing BC patients with healthy subjects, and reporting either OR or sufficient data to estimate the OR and its 95$\%$ CI. This results in inclusion of three new unbiased studies reporting OR. In this regard, we note that a  new study is included only if its sample of cases and controls do not overlap with the ones already included for meta-analysis. 

Following \citet{ruberu2023bayesian}, we include only those variants that are pathogenic for a study whenever that information was available. Specifically, if a study included information on both pathogenic variants and variants of uncertain significance (variants for which the pathogencity in unknown), we only count the number of pathogenic variants in computing the risk estimates. Moreover, when a study reported sufficient information, we tried to confirm whether a given variant classified as pathogenic by that particular study is still categorized as pathogenic by using ClinVar \citep{clinvar}, a free NIH archive of reports of human genetic variants. Our final ATM-BC meta-analysis is based on 30 studies that are summarized in Table \ref{asc.meta:ATM}.

\normalsize
  
 \subsubsection{Results}
\begin{figure}
     \centering
    \begin{subfigure}[b]{0.45\textwidth}
        \centering
        \includegraphics[width=\textwidth]{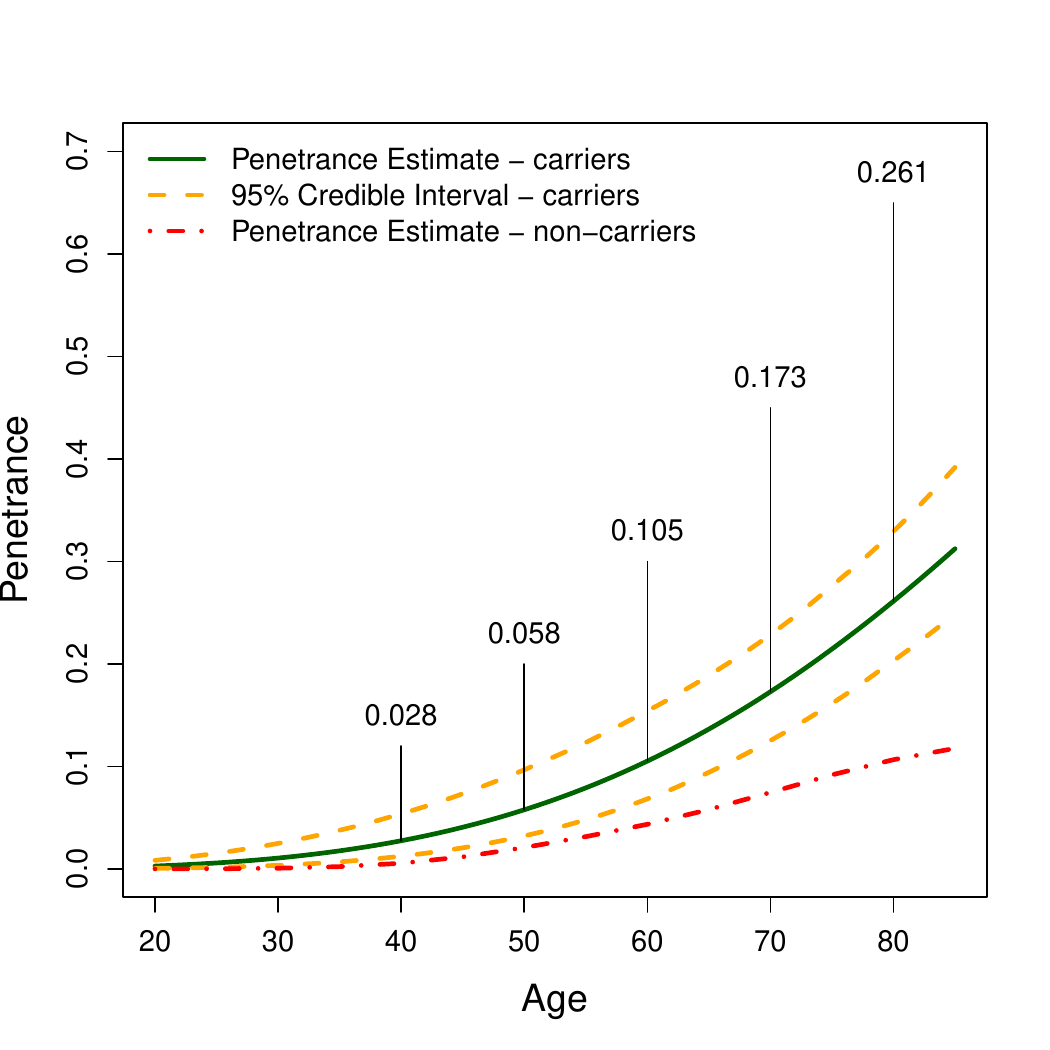}
        \caption{}
         \label{asc.ATM_penet.a}
     \end{subfigure}
     \hfill
     \begin{subfigure}[b]{0.45\textwidth}
         \centering
         \includegraphics[width=\textwidth]{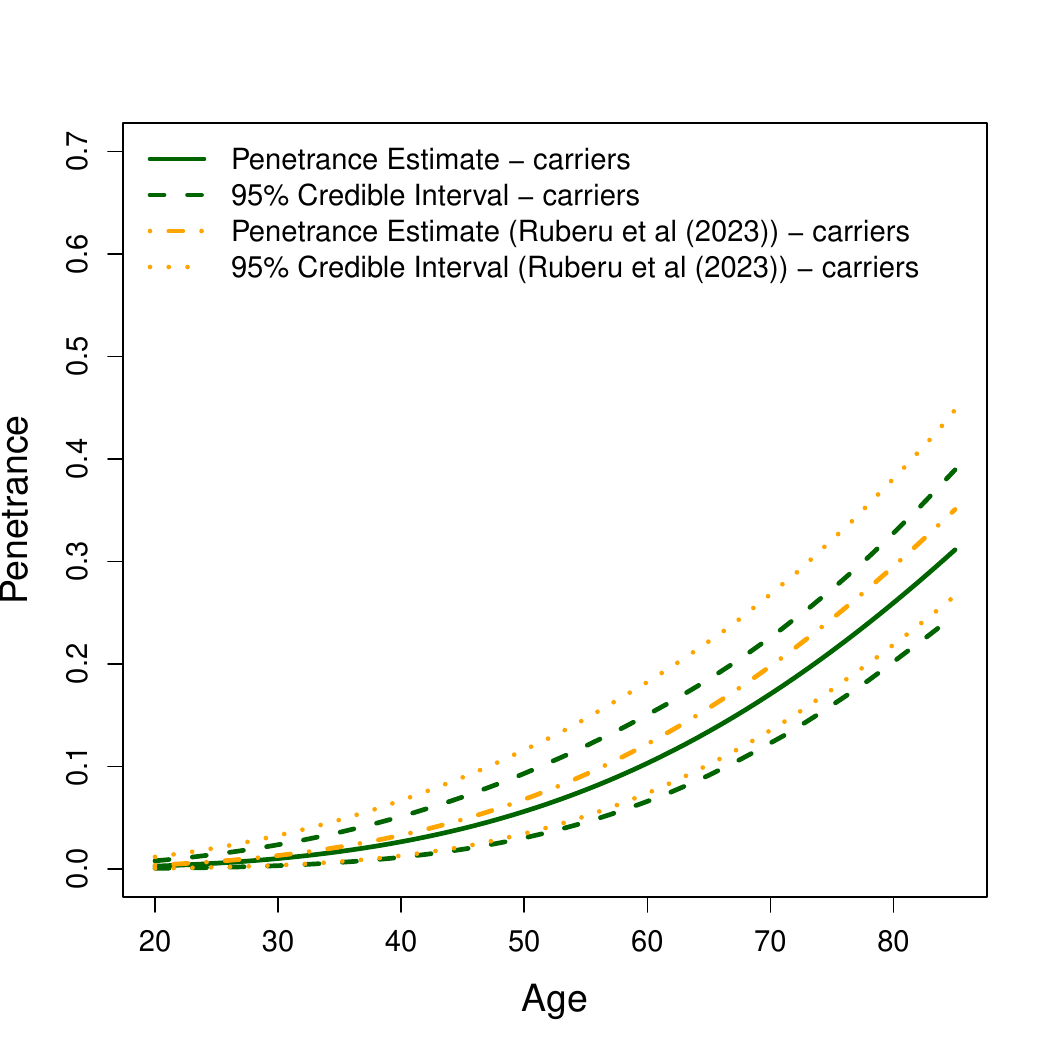}
         \caption{}
         \label{asc.ATM_penet.b}
     \end{subfigure}
     \caption{(a): Penetrance estimate for ATM-BC. (b): Comparison of the penetrance curve with that of \citet{ruberu2023bayesian}}
\end{figure}

 Figure \ref{asc.ATM_penet.a} shows the penetrance estimate of BC among individuals carrying an ATM pathogenic variant obtained using our proposed meta-analysis method. The risk of BC is $5.77\%$ ($3.22\%-9.67\%$) by age 50   and $26.13\%$ ($20.31\%-32.94\%$) by age 80. The penetrance estimates for non-carriers, as obtained from SEER data, are also shown in the figure. In Figure \ref{asc.ATM_penet.b}, we compare this penetrance curve with that of \citet{ruberu2023bayesian}, which did not include studies that may be biased due to ascertainment and the three new unbiased studies. Our estimated penetrance curve is lower than the one in \citet{ruberu2023bayesian}. We note that the credible intervals obtained by the proposed method are narrower showing the advantage of including more studies in the meta-analysis.

We also conduct a sensitivity analysis by removing studies one at a time that select cases based solely on age of onset \citep{fitzgerald1997heterozygous,brunet2008atm,dorling2021breast} or controls based on ages above 60 years \citep{momozawa2018germline}. The results are shown in Figure S3a. The resulting penetrance curves are very close to the overall curve indicating that our final curve is robust to inclusion of such studies



\subsection{Meta-Analysis of PALB2-BC Penetrance}

\subsubsection{Selection of Studies}
We start with all the unbiased studies used in \citet{ruberupalb2}, which were identified by implementing a semi-automated natural language processing–based procedure for abstract screening \citep{deng2019validation} and a Pubmed search of the following keywords in the title/abstract of the articles: "((((mutation) OR (variant)) AND (PALB2)) AND (breast cancer)) AND (germline)” up to 10 May 2023. These included 12 studies (2 studies reporting penetrance, 1 study reporting RR, and 9 studies reporting OR). Additionally, we include in our meta-analysis four studies subject to ascertainment bias (all reporting OR) making the total number of studies to be 16. As for ATM, we only included pathogenic variants of PALB2 for a study whenever that information was available. The 16 studies in our final PALB2-BC meta-analysis are summarized in Table \ref{asc.meta:PALB2}


\normalsize
\subsubsection{Results}
\begin{figure}
     \centering
    \begin{subfigure}[b]{0.45\textwidth}
        \centering
        \includegraphics[width=\textwidth]{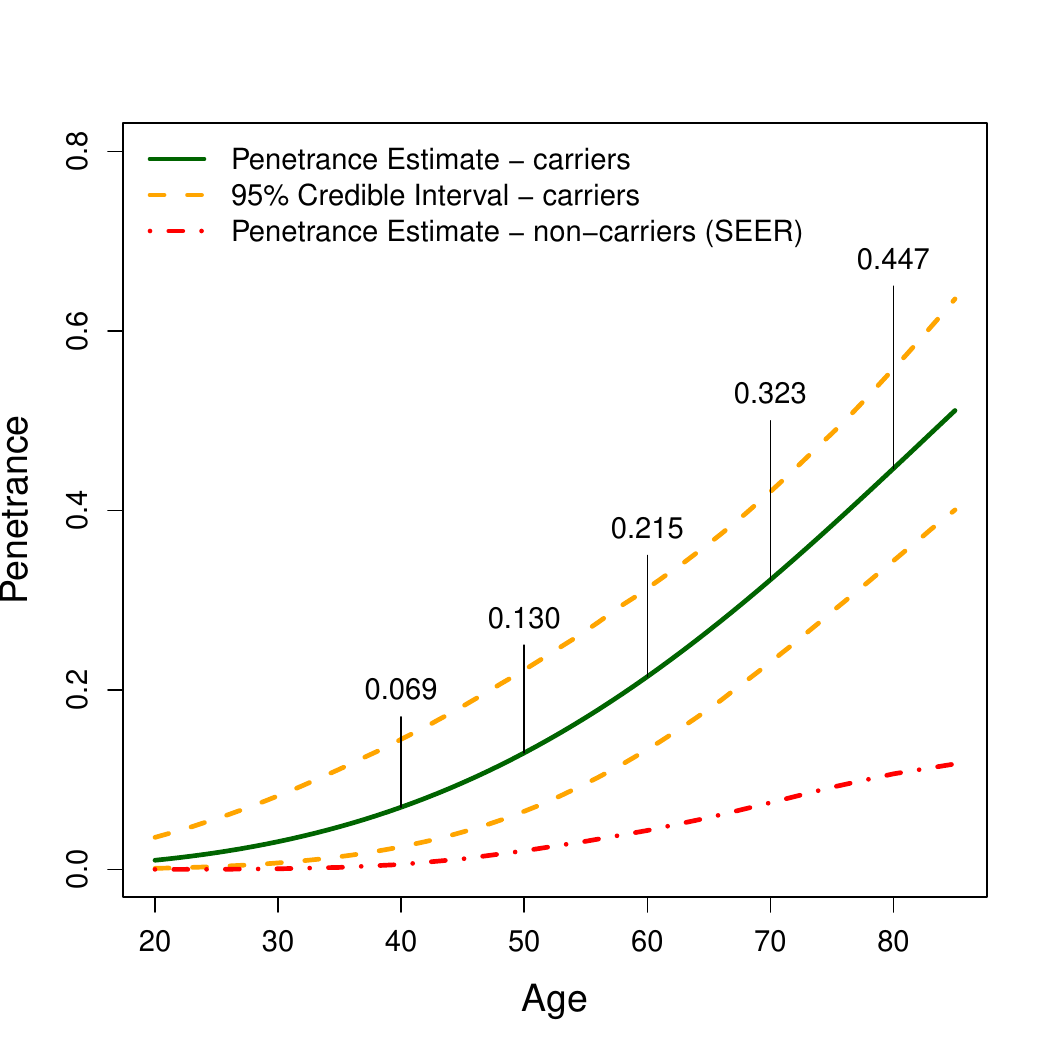}
        \caption{}
         \label{asc.PALB2.a}
     \end{subfigure}
     \hfill
     \begin{subfigure}[b]{0.45\textwidth}
         \centering
         \includegraphics[width=\textwidth]{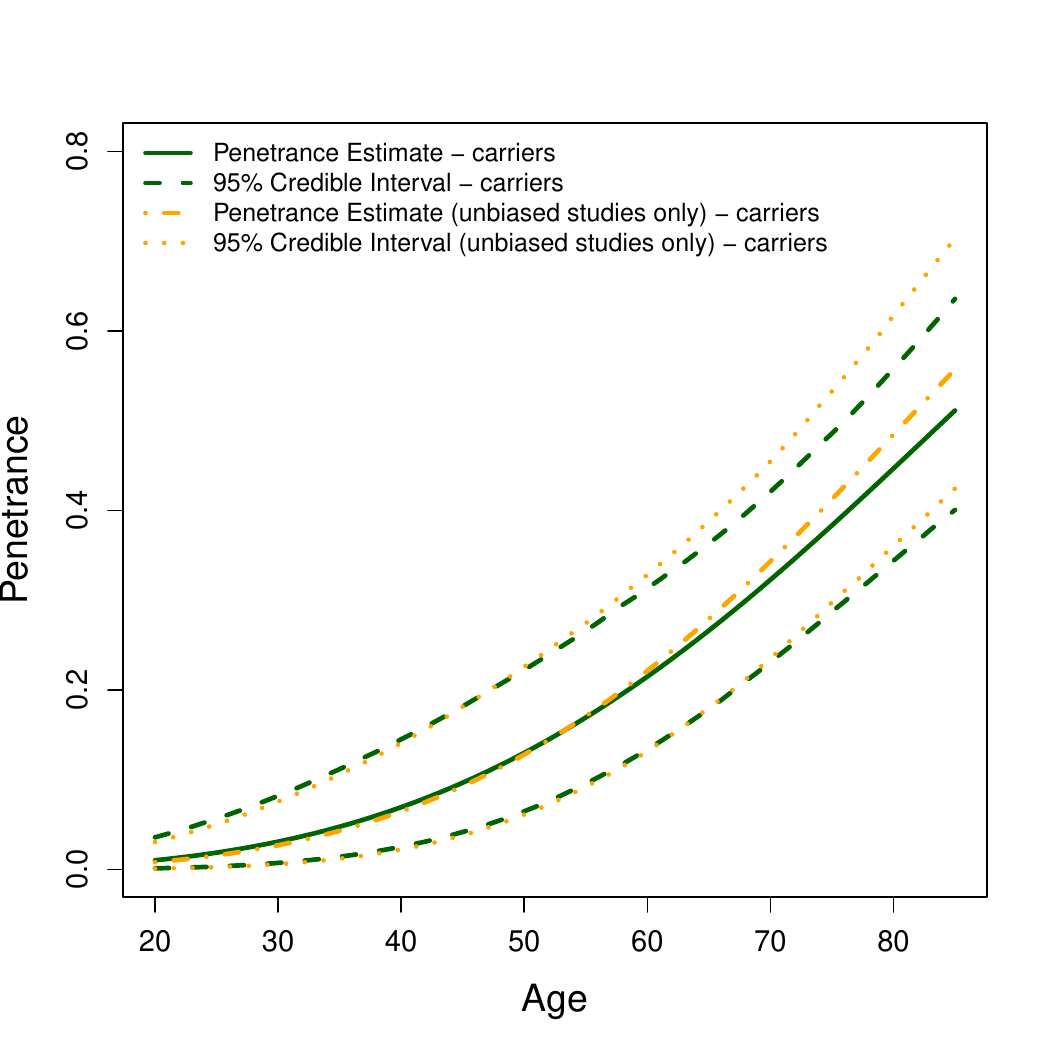}
         \caption{}
         \label{asc.PALB2.b}
     \end{subfigure}
     \caption{(a): Penetrance estimate for PALB2-BC. (b): Comparison of the penetrance curve with that of \citet{ruberu2023bayesian}}
\end{figure}

The penetrance curve of BC for carriers of pathogenic variants of the PALB2 gene is shown in Figure \ref{asc.PALB2.a} along with the penetrance curve for non-carriers (SEER). The risk of BC is $12.99\%$  ($6.48\%-22.23\%$) by age 50  and $44.69\%$ ($34.40\%-55.80\%$) by age 80. In Figure \ref{asc.PALB2.b}, we compare the penetrance curve with that of \citet{ruberupalb2} where studies subject to ascertainment bias were removed. The two curves overlap at ages $<$60 while after that the penetrance curve of \citet{ruberupalb2} is higher compared to the one from the proposed method. The 95$\%$ credible intervals of the two curves coincide until age 50. From age 50 onwards, the credible interval given by the proposed method is narrower.

As in the case of ATM, we conduct sensitivity analyses by removing studies that selected cases based on early onset \citep{cybulski2015clinical,dorling2021breast} or controls based on older ages \citep{momozawa2018germline}  one at a time. The resulting curves are very close to each other indicating the robustness of the overall penetrance estimates (Figure S3b).

\section{Discussion}

 The increasing popularity in multi-gene panel testing has resulted in a proliferation of studies examining the cancer risks associated with pathogenic variants in those genes \citep{Plichta2016}. The multi-gene panels encompass not only high-risk genes such as BRCA1/2 but also moderate-risk genes like ATM, CHEK2, and PALB2 \citep{turnbull2008genetic}. As expected, the risk estimates 
 for a specific gene vary across studies due to not only sampling variability but also variations in research study design  and the type of reported results (age-specific penetrance, OR, RR, and SIR). Moreover, not all studies provide unbiased risk estimates. In observational case-control studies, ascertainment
bias can occur when cases and controls are ascertained differently (cases based on family history and controls from the general population). This poses a challenge because inclusion of such studies in meta-analysis without bias adjustment can lead to biased estimates. Thus, typically such studies are excluded as was done by \citet{ruberu2023bayesian}. In that paper, the authors had proposed a novel Bayesian hierarchical random-effects model that can integrate information available from different types of studies to provide age-specific penetrance. Here, we propose an extension of that method to allow the inclusion of such studies in meta-analysis by adjusting for the ascertainment bias via the model. This allows for the inclusion of biased studies that can provide useful information about risk if the bias is adjusted for in the meta-analysis. 
 
 Our simulation study shows that including biased OR estimates from case-control studies in meta-analysis (Settings 1 through 3) with correction for the bias by applying our proposed method generally leads to lower RMSE/True values compared to excluding these studies altogether. Simulation results further indicate that the coverage probabilities of $95\%$ CrI remain consistent at approximately $95\%$ under the proposed method. Our simulation study also highlights that including biased studies in the analysis without corrections leads to overestimated penetrance values, higher RMSE/True values, and decreased coverage probabilities of $95\%$ CrI, especially at ages 70 and 80. We made similar observations in Setting 4 with the exception of having a higher RMSE/True values at age 40 with bias correction compared to other two approaches. Recall that, in Setting 4, we added hypothetical five biased RR studies into the meta-analysis (in addition to the ten OR biased studies) even though we did not find any such studies in our literature search; thus, this setting is of less practical interest. Altogether, these results provide evidence that our model performs well in a wide variety of settings. 
 
 We also provide updated penetrance estimates for BC due to pathogenic variants of ATM and PALB2  by including studies with ascertainment bias in meta-analysis. As these updated estimates are based on larger number of studies, they are likely to be more robust than the ones provided in \citet{ruberu2023bayesian} and \citet{ruberupalb2}. They are also more efficient as reflected in narrower 95$\%$ confidence intervals.

Finally, we acknowledge some limitations of our study. Recall that for both ATM and PALB2, only two studies reported age-specific penetrance. None of these four studies provide penetrance values for ages below 35. This may be because BC is rare at younger ages. Furthermore, none of the OR/RR estimates included in the analysis are age-specific. Thus, due to lack of direct empirical data, caution should be exercised in interpreting the reported meta-analysis penetrance curves for ages under 35. In addition, we ignored potential ascertainment bias in studies that selected cases based solely on age of onset but not family history.  However, our sensitivity analysis removing such studies one at a time from both the meta-analyses indicate that our results are robust to inclusion of these studies (Figure  S3). It is also noteworthy that, due to lack of studies quantifying ascertainment bias in case-control studies, we rely on bias estimates from RCTs to establish the prior distributions for our bias parameters. Moreover, we focus here solely on mitigating ascertainment bias. There may be other types of bias such as confounding bias, publication bias, and recall bias \citep{hemkens2018interpretation, harewood2005assessment, coughlin1990recall}. Future research could explore adjustments for these biases within the framework of meta-analysis. Despite these limitations, we believe that the proposed method advances the field of meta-analysis of observational studies towards a new direction by allowing inclusion of studies subject to ascertainment bias and providing more robust and efficient penetrance estimates.

\section{Software}

An R package BayesMetaPenetrance version 1.1 implementing the proposed method is under construction and will be available at \url{https://personal.utdallas.edu/~sxb125731/} and \url{https://github.com/LakshikaRuberu}.

\section*{Acknowledgements}
This work is supported by NIH grant R03CA242562-01.

\bibliography{references}
\clearpage

\newpage
\scriptsize
\begin{spacing}{.7}
\begin{longtable} {llcccc}  
 \caption{Summary of studies included in the meta-analysis of ATM-BC penetrance}\\
      \label{asc.meta:ATM}
 Index &\makecell{Study} & \makecell{Cases} & \makecell {Study Design} & \makecell{Sample Size} & \makecell{Risk and CI\\ (Input for \\the methods)}\\\hline \hline 

            1& \cite{goldgar2011rare} &\makecell{ Familial BC\\ BRCA1/2 negative}&\makecell{Family-based \\segregation analysis}&156 &\makecell{Penetrance\\ Curve}  \\ 

            2& \cite{thompson2005cancer} & \makecell{One family
member \\with ATM}&\makecell{Cancer incidence in \\relatives of ATM
patients} &1160&\makecell{Penetrance\\ Curve} \\

            3& \cite{swift2008breast} &\makecell{One family
member \\with ATM}& \makecell{Cancer incidence in \\relatives of ATM
patients}  & 919&\makecell{RR = 2.4\\
(1.3 - 4.3) } \\

            4& \cite{renwick2006atm} & \makecell{ Familial BC\\ BRCA1/2 negative}& \makecell{Case-control/family-based
\\segregation analysis}  & 5173&\makecell{RR = 2.37\\(1.51 - 3.78)}\\ 

5& \cite{Li2016} & \makecell{Familial BC\\ BRCA1/2 negative} &\makecell{family-based
\\segregation analysis}& 660&\makecell{RR = 2.67\\(0.82 - 10.56)}\\
             
             6&\cite{olsen2005breast} & \makecell{One family member \\with ATM} &\makecell{Cancer incidence in \\relatives of ATM
patients}& 712&\makecell{SIR = 2.9 \\ (1.9 - 4.4)} \\ 
             
            7& \cite{andrieu2005ataxia} & \makecell{One family member \\with ATM} &\makecell{Cancer incidence in \\relatives of ATM
patients}& 708&\makecell{SIR = 2.43\\(1.32 - 4.09)}\\

8&  \cite{Kurian2017} & \makecell{Female patients who \\underwent panel testing} &\makecell{Case control study}&95561&\makecell{OR = 1.74 \\(1.46 - 2.07)}\\

9& \cite{momozawa2018germline} & \makecell{Unselected BC } & \makecell{Case-control study}  & 18292&\makecell{OR = 2.10\\(1.0 - 4.1)}\\ [2pt]  
            
            10&\cite{dorling2021breast} &  \makecell{Mainly unselected BC with\\ a subset of early onset BC \\ BRCA1/2 negative} & \makecell{Case-control study} &97997& \makecell{OR = 2.10$^a$\\ (1.71 - 2.57)} \\

        11& \cite{hu2021population}  & \makecell{Unselected BC } & \makecell{Case-control study} & 64791 & \makecell{OR = 1.82$^{b,d}$\\(1.46 - 2.27)}\\

             12& \cite{mangone2015atm}  & \makecell{Sporadic BC} & \makecell{Case-control study} & 200& \makecell{OR = 3.03$^{c,d,e}$\\(NA)} \\

             13& \cite{brunet2008atm}  & \makecell{Unselected early-onset \\BC ($<46$ years)\\
BRCA1/2 negative} &\makecell{Case-control study} & 193&\makecell{OR = 18.13$^{c,d}$ \\ (NA)}\\

            14& \cite{pylkas2007evaluation}  & \makecell{Familial and \\ unselected BC }  & \makecell{Case-control study}&2231&\makecell{OR = 6.93$^{e,f}$\\ (0.85 - 56.43)}\\ 
            
              15& \cite{Zheng2018}  & \makecell{Unselected BC} &\makecell{Case-control study}&2133&\makecell{OR = 4.40$^{e}$\\(0.51 - 37.75)} \\ 
             
             16& \cite{kreiss2000founder}  & \makecell{Unselected BC} &\makecell{Case-control study}&298&\makecell{OR = 3.09$^{e}$\\(0.50 - 18.96)} \\ 
             
            17&  \cite{fitzgerald1997heterozygous}& \makecell{Early-onset BC \\($<40$ years)} &\makecell{Case-control study}&603&\makecell{OR = 0.50$^{e}$\\(0.07 - 3.58)}\\ 

18& \cite{ahearn2022breast} & \makecell{Recommended for biopsy or\\documented BC} & \makecell{Case-control study}  & 2434&\makecell{OR$^{h}$ = 1.60\\(0.42-6.10)}\\ 

             19&  \cite{nurmi2022pathogenic}& \makecell{Unselected BC} &\makecell{Case-control study}&2468&\makecell{OR = 2.10$^{e,f,h}$\\(0.40 - 10.61)}\\

20&  \cite{felix2022mutational}& \makecell{Familial BC} &\makecell{Case-control study}&290&\makecell{OR = 4.96$^{c,e,h}$\\(NA)}\\
                
                21&  \cite{hauke2018gene}& \makecell{Familial or early-onset BC\\ ($<$36 years; if
bilateral $<$51 years)\\BRCA1/2 negative} &\makecell{Case-control study}&7778&\makecell{OR = 3.12$^{g}$\\(1.56 - 6.25)}\\

           22&  \cite{thompson2016panel}& \makecell{Familial BC \\BRCA1/2 negative} &\makecell{Case-control study}&3978&\makecell{OR = 2.15$^{g}$\\(0.69 - 7.33)}\\

            23&  \cite{couch2017associations}& \makecell{Familial BC} &\makecell{Case-control study}&93314&\makecell{OR = 2.15$^{g}$\\(2.41 - 3.50)}\\

             24&  \cite{girard2019familial}& \makecell{Familial BC\\BRCA1/2 negative} &\makecell{Case-control study}&2406&\makecell{OR = 1.80$^{g}$\\(1.20 - 2.70)}\\

            25&  \cite{grana2011germline}& \makecell{Familial BC\\ BRCA1/2 negative} &\makecell{Case-control study}&948&\makecell{OR = 5.76$^{c,e,g}$\\(NA)}\\

            26&  \cite{allinen2002atm}& \makecell{Familial and sporadic BC\\ BRCA1/2 negative} &\makecell{Case-control study}&446&\makecell{OR = 5.82$^{c,e,g}$\\(NA)}\\

             27&  \cite{thorstenson2003contributions}& \makecell{Familial BC} &\makecell{Case-control study}&322&\makecell{OR = 1.79$^{c,d,e,g}$\\(NA)}\\
            
            28&  \cite{soukupova2008contribution}& \makecell{Familial or early-onset BC\\ ($<$40 years; if
bilateral $<$50 years)\\ BRCA1/2 negative} &\makecell{Case-control study}&344&\makecell{OR = 8.10$^{c,d,e,g}$\\(NA)}\\

            29&  \cite{teraoka2001increased}& \makecell{family
history of BC and/or \\Early-onset BC ($<$35 years)   } &\makecell{Case-control study}&223&\makecell{OR = 1.73$^{c,d,e,g}$\\(NA)}\\

            30&  \cite{izatt1999identification}& \makecell{
moderate/absent family history\\Early-onset BC ($<$40 years)\\BRCA1/2 negative} &\makecell{Case-control study}&206&\makecell{OR = 3.21$^{c,d,e,g}$\\(NA)}\\

            \hline
\multicolumn{6}{l}{\scriptsize $^a$ Based on 30 studies in breast cancer association consortium (BCAC) unselected for family history.} \\ 
 
 \multicolumn{6}{l}{\scriptsize $^b$ Based on 12 studies in the CARRIERS
consortium not enriched with patients
with a family history or early onset of disease.}\\

\multicolumn{6}{l}{\scriptsize $^c$ No mutations in controls, $^d$ Excluded VUSs, $^e$  OR and CI were calculated using data reported in the paper.}\\
\multicolumn{6}{l}{\scriptsize $^f$  Reported number of mutations for both familial and unselected cases. Only unselected cases were included.}\\
\multicolumn{6}{l}{\scriptsize $^g$  Subjected to ascertainment bias }\\ 
\multicolumn{6}{l}{\scriptsize $^h$  Unbiased studies not included in \citet{ruberu2023bayesian} meta-analysis}\\ 

         \end{longtable}
\end{spacing}

\newpage

\scriptsize
\begin{spacing}{.7}
\begin{longtable} {llcccc}  
 \caption{Summary of studies included in the meta-analysis of PALB2-BC penetrance}\\
      \label{asc.meta:PALB2} 
 
 Index &\makecell{Study} & \makecell{Cases} & \makecell {Study Design} & \makecell{Sample Size} & \makecell{Risk and CI\\ (Input to \\meta-analysis)}\\\hline \hline 

1& \cite{Antoniou2014} &\makecell{One family
member \\with PALB2\\ BRCA1/2 negative}&\makecell{Family-based \\segregation analysis}&311 &\makecell{Penetrance\\ Curve}  \\ 

2& \cite{Erkko2008} & \makecell{Family of PALB2 carrier \\with BC}&\makecell{Family-based \\segregation analysis} &213&\makecell{Penetrance\\ Curve} \\

3& \cite{rahman2007palb2} &\makecell{ Familial BC \\BRCA1/2 negative}& \makecell{Case-control/family-based
\\segregation analysis}  & 2007&\makecell{RR = 2.3\\
(1.4 - 3.9) } \\

4&  \cite{Kurian2017} & \makecell{Female patients who \\underwent panel testing} &\makecell{Case control study}&95561&\makecell{OR = 3.39 \\(2.79 - 4.12)}\\

5& \cite{momozawa2018germline} & \makecell{Unselected BC } & \makecell{Case-control study}  & 18292&\makecell{OR = 9.00\\(3.4-29.7)}\\  
            
6&\cite{dorling2021breast} &  \makecell{Mainly unselected BC with\\ a subset of early onset BC \\ BRCA1/2 negative} & \makecell{Case-control study} &97997& \makecell{OR = 5.02$^a$\\ (3.73-6.76)} \\ 

7& \cite{cybulski2015clinical} & \makecell{Recruited from hospitals with \\ a subset of early onset BC \\($<$50 years) } & \makecell{Case-control study}  & 17231&\makecell{OR = 4.39\\(2.30-8.37)}\\ 

8& \cite{heikkinen2009breast} & \makecell{Sporadic BC} & \makecell{Case-control study}  & 2353&\makecell{OR = 3.4$^{d}$\\(0.68-32.95)}\\ 
9& \cite{ahearn2022breast} & \makecell{Recommended for biopsy or\\documented BC} & \makecell{Case-control study}  & 2434&\makecell{OR = 17.25\\(2.15-138.13)}\\ 
            
10& \cite{Zheng2018}  & \makecell{Unselected BC} &\makecell{Case-control study}&2133&\makecell{OR = 20.384$^{b,c}$\\(NA)} \\ 

11& \cite{diaz2022evaluating}  & \makecell{Unselected BC} &\makecell{Case-control study}&3286&\makecell{OR = 3.76\\(0.66 - 21.53)} \\

12&  \cite{felix2022mutational}& \makecell{Familial BC} &\makecell{Case-control study}&290&\makecell{OR = 4.96$^{b,c}$\\(NA)}\\

13&  \cite{hauke2018gene}& \makecell{Familial or early-onset BC\\ ($<$36 years; if
bilateral $<$51 years)\\BRCA1/2 negative} &\makecell{Case-control study}&7778&\makecell{OR = 12.67$^{e}$\\(3.10 - 51.79)}\\

14&  \cite{thompson2016panel}& \makecell{Familial BC \\BRCA1/2 negative} &\makecell{Case-control study}&3978&\makecell{OR =6.56$^{e}$\\(2.29 - 18.8)}\\

15&  \cite{couch2017associations}& \makecell{Familial BC} &\makecell{Case-control study}&94964&\makecell{OR = 6.25$^{e}$\\(4.82 - 8.14)}\\

16&  \cite{girard2019familial}& \makecell{Familial BC\\BRCA1/2 negative} &\makecell{Case-control study}&2406&\makecell{OR = 3.20$^{e}$\\(1.50 - 6.90)}\\

\hline
 \multicolumn{6}{l}{\scriptsize $^a$ Based on 30 studies in breast cancer association consortium (BCAC) unselected for family history.} \\ 
 \multicolumn{6}{l}{\scriptsize $^b$ No mutations in controls}\\
\multicolumn{6}{l}{\scriptsize $^c$  OR and CI were calculated using data reported in the paper}\\
\multicolumn{6}{l}{\scriptsize $^d$  Only Sporadic cases were considered}\\
\multicolumn{6}{l}{\scriptsize $^e$  Subjected to ascertainment bias}
            \end{longtable}
            \end{spacing}
            
\begin{table}[htbp!]
\centering
\caption{Results for Setting 1 (30 studies with 10 biased OR) -- Scenario 1}
\label{Table1}
\begin{tabular}{ccccccc}\hline
& & \multicolumn{5}{c}{Ages}\\
& & 40 & 50 & 60 & 70 & 80\\ \hline
\multirow{4}{*}{\makecell{Penetrance\\ Estimates}}&True Penetrance& 0.026 &	0.067&	0.141&	0.253&	0.398\\\cline{2-6}
&Approach 1& 0.028    &    0.068 &       0.139  &    0.250 &       0.400 \\
&Approach 2&  0.030    &    0.074  &      0.156 &       0.286 &       0.459  \\
&Approach 3& 0.029&	0.070&	0.142&	0.254&	0.406\\
\hline
\multirow{3}{*}{$\cfrac{\text{RMSE}}{\text{True}}$}
&Approach 1& 0.3196& 0.2318 &0.1766& 0.1390& 0.1125   \\
&Approach 2& 0.3608 &0.2666 &0.2105& 0.1825 &0.1777 \\ 
&Approach 3& 0.3531	&0.2562& 0.1938	&0.1509 &0.1219\\
\hline
\multirow{3}{*}{95\% CrI coverage}
&Approach 1& 0.986& 0.986 &0.974& 0.960& 0.930   \\
&Approach 2& 0.986& 0.986 &0.974 &0.916& 0.784   \\
&Approach 3& 0.990 &0.992& 0.980& 0.964& 0.930 \\
\hline
\end{tabular}
\end{table}

\begin{table}[htbp!]
\centering
\caption{Results for Setting 2 (30 studies with 5 biased OR) -- Scenario 1}
\label{Table2}
\begin{tabular}{ccccccc}\hline
& & \multicolumn{5}{c}{Ages}\\
& & 40 & 50 & 60 & 70 & 80\\ \hline
\multirow{4}{*}{\makecell{Penetrance\\ Estimates}}&True Penetrance& 0.026 &	0.067&	0.141&	0.253&	0.398\\\cline{2-6}
&Approach 1& 0.028 &       0.068  &      0.139   &     0.250  &      0.400  \\
&Approach 2&  0.028      &  0.070     &   0.149   &     0.274  &      0.442  \\
&Approach 3& 0.029&	0.070&	0.142&	0.254&	0.406\\
\hline
\multirow{3}{*}{$\cfrac{\text{RMSE}}{\text{True}}$}
&Approach 1& 0.3098 &0.2285 &0.1738& 0.1365 &0.1110  \\
&Approach 2& 0.3335 &0.2479& 0.1909 &0.1581& 0.1493   \\ 
&Approach 3& 0.3531	&0.2562& 0.1938	&0.1509 &0.1219\\
\hline
\multirow{3}{*}{95\% CrI coverage}
&Approach 1& 0.992 &0.992& 0.978 &0.960 &0.934    \\
&Approach 2& 0.986 &0.986 &0.980& 0.956& 0.864   \\
&Approach 3& 0.990 &0.992& 0.980& 0.964& 0.930 \\
\hline
\end{tabular}
\end{table} 

\begin{table}[htbp!]
\centering
\caption{Results for Setting 3 (27 studies with 15 biased OR) -- Scenario 1}
\label{Table3}
\begin{tabular}{ccccccc}\hline
& & \multicolumn{5}{c}{Ages}\\
& & 40 & 50 & 60 & 70 & 80\\ \hline
\multirow{4}{*}{\makecell{Penetrance\\ Estimates}}&True Penetrance& 0.026 &	0.067&	0.141&	0.253&	0.398\\\cline{2-6}
&Approach 1&   0.029  &      0.069    &    0.141 &       0.251  &      0.399   \\
&Approach 2&  0.033   &     0.083     &   0.175  &      0.318   &     0.504 \\
&Approach 3& 0.029&	0.070&	0.142&	0.254&	0.406\\
\hline
\multirow{3}{*}{$\cfrac{\text{RMSE}}{\text{True}}$}
&Approach 1& 0.3300& 0.2379 &0.1803 &0.1413& 0.1134   \\
&Approach 2& 0.4525& 0.3493& 0.2931& 0.2758 &0.2739    \\ 
&Approach 3& 0.3531	&0.2562& 0.1938	&0.1509 &0.1219\\
\hline
\multirow{3}{*}{95\% CrI coverage}
&Approach 1& 0.992& 0.990 &0.986 &0.966& 0.942  \\
&Approach 2&0.986& 0.976 &0.932& 0.794& 0.566   \\
&Approach 3& 0.990 &0.992& 0.980& 0.964& 0.930 \\
\hline
\end{tabular}
\end{table} 

\clearpage

\clearpage

\end{document}


\begin{center}
\bf{\large{Supplementary Materials for \\ Adjusting for Ascertainment Bias in Meta-Analysis of Penetrance for Cancer Risk}}
\end{center}

\section*{Details of the MCMC Algorithm}

At iteration $(t+1)$ of the algorithm, the following updates are made.$\\$
{\bf {Update of $\mathbf{\kappa_s}$ (Unbiased Studies):}} The full conditional distribution of $\kappa_s$ is given by 
\[ \pi(\kappa_s|y_s,w_s^*, \lambda_s, a, b ) \propto L(\btheta_s)\pi(\kappa_s|a, b). \]
We use a Gamma$(\alpha_s, \beta_s)$ proposal whose mean $\alpha_s\beta_s$ is set to be equal to $\kappa_s^{(t)}$ (the value of $\kappa_s$ at previous iteration) and variance $\alpha_s\beta_s^2$ is set to be equal to a function of the current value of $\kappa_s^{(t)}$. In particular, for studies reporting age-specific penetrance, $\alpha_s\beta_s^2 =$ log$_{500}(\kappa_s^{(t))}+0.01)$, for studies reporting RR, it is log$_{500}(\kappa_s^{(t))}+2000)$ and for those reporting OR, it is log$_{500}(\kappa_s^{(t))}+200000)$. These choices ensure convergence and recommended acceptance rates (Gelman et al., 2013, Chapters~11-12).$\\$
{\bf {Update of $\mathbf{\lambda_s}$ (Unbiased Studies):}} $\lambda_s$ is updated in a similar way as $\kappa_s$
using a Gamma proposal whose mean is $ \lambda_s^{(t)}$ and variance is set to be equal to $\lambda_s^{(t)^{0.4}}, \lambda_s^{(t)^{0.9}},$ and $\lambda_s^{(t)^{1.2}}$ for studies reporting age-specific penetrance, RR, and OR, respectively.  $\\$
{\bf {Update of $\mathbf{\kappa_s}$ and $\mathbf{\lambda_s}$ (Studies with Ascertainment Bias):}} The full conditional distribution of $\kappa_s$ is given by 
\begin{align*}
\pi(\kappa_s|y_s, w_s^*,\lambda_s, B_s, a, b) &\propto L(\btheta_s, B_s)\pi(\kappa_s|a, b)\\
& \propto \frac{1}{\sqrt{2\pi w_s^*}} \exp{\left(\frac{-1}{2} \bigg(\frac{log(y_s)-(log(\nu_s)+B_s)}{\sqrt{w_s^*}}\bigg)^2\right)} \times\\& \hspace{3in}\left(\frac{1}{\Gamma(a) b^{a}}\right)(\kappa_s)^{a-1}\exp{\left(-\kappa_s/b\right)}\\
& \propto \exp{\left(\frac{-1}{2} \bigg(\frac{log(y_s)-(log(\nu_s)+B_s)}{\sqrt{w_s^*}}\bigg)^2\right)}(\kappa_s)^{a-1}\exp{\left(-\kappa_s/b\right),}
\end{align*}
where $w_s^*$ is the variance of $\log(y_s)$ and $\nu_s$ is $y_s$ expressed in terms of penetrance parameters $\kappa_s$, and $\lambda_s$.  $B_s$ is $B_s^{RR}$ or $B_s^{OR}$. $\lambda_s$ is updated in a similar way. We utilize the same proposal distributions as described earlier for $\kappa_s$, and $\lambda_s$, depending on the type of study being analyzed.$\\$
{\bf {Update of $\mathbf{B_s}$ (Studies with Ascertainment Bias):}} The full conditional distribution of $B_s$ is given by
\begin{align*}
\pi(B_s|y_s, w_s^*,\kappa_s, \lambda_s, a, b) &\propto L(\btheta_s, B_s)\pi(B_s|\sigma)\\
& \propto \frac{1}{\sqrt{2\pi w_s^*}} \exp{\left(\frac{-1}{2} \bigg(\frac{log(y_s)-(log(\nu_s)+B_s)}{\sqrt{w_s^*}}\bigg)^2\right)} \frac{\sqrt{2}}{\sqrt{\pi}\sigma} \exp{\left(\frac{-1}{2}\bigg(\frac{B_s}{\sigma}\bigg)^2\right)}\\
& \propto \exp{\left(\frac{-1}{2} \bigg(\frac{log(y_s)-(log(\nu_s)+B_s)}{\sqrt{w_s^*}}\bigg)^2\right)} \exp{\left(\frac{-1}{2}\bigg(\frac{B_s}{\sigma}\bigg)^2\right).}
\end{align*}
 The proposal distribution for $B_s$ is half-normal  with $\sigma=0.836 B_s^{(t)}$.
$\\$
{\bf {Update of $a$:}} The conditional distribution of $a$ is given by 
\[\pi(a|\by_s, \bkappa, b)  \propto \pi(\bkappa|a, b)\pi(a). \]
We use a uniform proposal distribution U$(\max(l_a, a^{(t)} - 9), \; \min(a^{(t)} + 9, u_a))$.$\\$
$\\$
{\bf {Update of $b$:}} The conditional distribution of $b$ is given by 
\[\pi(b|\by_s, \bkappa, a)  \propto \pi(\bkappa|a, b)\pi(b). \]
We use a uniform proposal distribution U$(\max(l_b, b^{(t)} - 0.04), \; \min(b^{(t)} + 0.04, u_b))$.$\\$
$\\$
{\bf {Update of $c$ and $d$:}} These are carried out in a similar manner as for $a$ and $b$ with uniform proposal distributions of U$(\max(l_c, c^{(t)} - 8), \; \min(c^{(t)} + 8, u_c))$ and $\\$U$(\max(l_d, d^{(t)} - 0.22), \; \min(d^{(t)} + 0.22, u_d))$, respectively.

\newpage

\begin{table}[htbp!]
\centering
\caption{Results for Setting 4 (35 studies with 10 biased OR, and 5 biased RR) -- Scenario 1}
\label{Table4}
\begin{tabular}{ccccccc}\hline
& & \multicolumn{5}{c}{Ages}\\
& & 40 & 50 & 60 & 70 & 80\\ \hline
\multirow{4}{*}{\makecell{Penetrance\\ Estimates}}
&True Penetrance& 0.026 &	0.067&	0.141&	0.253&	0.398\\\cline{2-6}
&Approach 1&  0.030     &     0.071   &       0.143   &       0.252  &        0.399   \\
&Approach 2&    0.028  &      0.073 &       0.157   &     0.292  &      0.472 \\
&Approach 3& 0.029&	0.070&	0.142&	0.254&	0.406\\
\hline
\multirow{3}{*}{$\cfrac{\text{RMSE}}{\text{True}}$}
&Approach 1& 0.3722 &0.2568 &0.1852& 0.1387 &0.1083  \\
&Approach 2& 0.3505 &0.2642 &0.2114& 0.1913& 0.1988   \\
&Approach 3& 0.3531	&0.2562& 0.1938	&0.1509 &0.1219\\
\hline
\multirow{3}{*}{95\% CrI coverage}
&Approach 1& 0.984 &0.980 &0.970& 0.948& 0.938   \\
&Approach 2&0.974 &0.970 &0.962& 0.904 &0.710   \\
&Approach 3& 0.990 &0.992& 0.980& 0.964& 0.930 \\
\hline
\end{tabular}
\end{table}

\begin{table}[htbp!]
\centering
\caption{Results for Setting 1 (30 studies with 10 biased OR) -- Scenario 2}
\label{Table5}
\begin{tabular}{ccccccc}\hline
& & \multicolumn{5}{c}{Ages}\\
& & 40 & 50 & 60 & 70 & 80\\ \hline
\multirow{4}{*}{\makecell{Penetrance\\ Estimates}}&True Penetrance& 0.026 &	0.067&	0.141&	0.253&	0.398\\\cline{2-6}

&Approach 1& 0.026  &      0.065  &      0.136 &       0.247  &      0.399 \\
&Approach 2&  0.027  &     0.070  &      0.151  &      0.281  &      0.457 \\
&Approach 3& 0.027&	0.066&	0.137&	0.250&	0.403\\
\hline
\multirow{3}{*}{$\cfrac{\text{RMSE}}{\text{True}}$}

&Approach 1& 0.2780& 0.2058 &0.1569 &0.1240&0.1033 \\
&Approach 2&0.2801 &0.2111& 0.1698 &0.1584 &0.1695\\
&Approach 3& 0.2855 &0.2123& 0.1648 &0.1318 &0.1107\\
\hline
\multirow{3}{*}{95\% CrI coverage}
&Approach 1& 0.992& 0.986& 0.982& 0.960 &0.946  \\
&Approach 2& 0.992 &0.988& 0.984& 0.934 &0.788 \\
&Approach 3& 0.998& 0.994& 0.988& 0.966& 0.944 \\
\hline
\end{tabular}
\end{table}

\begin{table}[htbp!]
\centering
\caption{Results for Sensitivity Analysis (30 studies with 10 biased OR) - Scenario 1}
\label{Table6}
\begin{tabular}{ccccccc}\hline
& & \multicolumn{5}{c}{Ages}\\
& & 40 & 50 & 60 & 70 & 80\\ \hline
\multirow{3}{*}{\makecell{Penetrance\\ Estimates}}
&True Penetrance& 0.026 &	0.067&	0.141&	0.253&	0.398\\
&half-normal $(\sigma=1.09)$& 0.028    &    0.066    &    0.136    &    0.245  &      0.393 \\
&half-normal $(\sigma=0.76)$&   0.029   &     0.069      &  0.142   &     0.254  &      0.406 \\
\hline
\multirow{2}{*}{$\cfrac{\text{RMSE}}{\text{True}}$}
&half-normal $(\sigma=1.09)$& 0.3129 &0.2296& 0.1757 &0.1388 &0.1118  \\
&half-normal $(\sigma=0.76)$& 0.3282& 0.2365 &0.1779& 0.1387& 0.1137 \\
\hline
\multirow{2}{*}{95\% CrI coverage}
&half-normal $(\sigma=1.09)$& 0.986& 0.988 &0.972 &0.946 &0.932  \\
&half-normal $(\sigma=0.76)$&0.988 &0.986 &0.984& 0.964 &0.934 \\
\hline
\end{tabular}
\end{table}

  \begin{figure}
     \centering
    \begin{subfigure}[b]{0.48\textwidth}
        \centering
        \includegraphics[width=\textwidth]{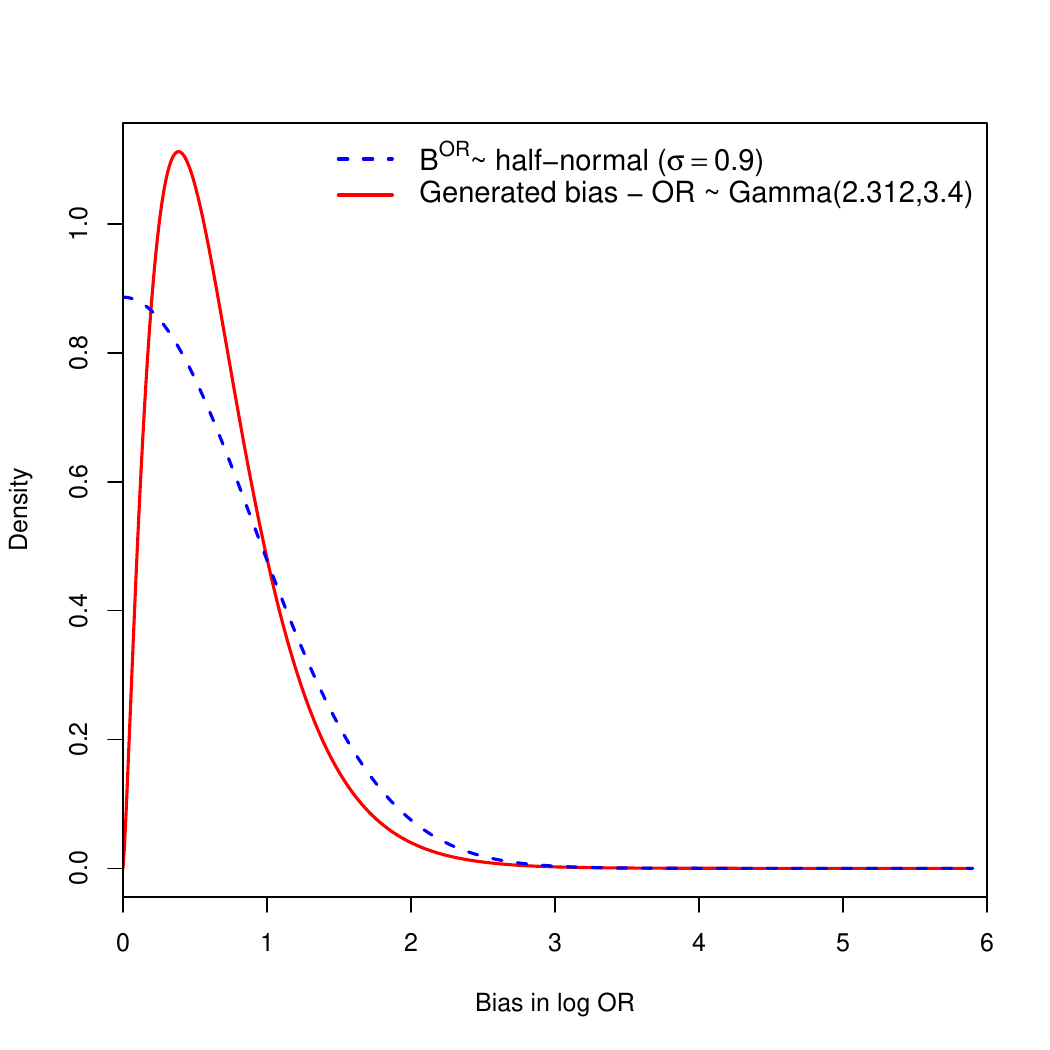}
        \caption{}
         \label{asc.prior.a}
     \end{subfigure}
     \hfill
     \begin{subfigure}[b]{0.48\textwidth}
         \centering
         \includegraphics[width=\textwidth]{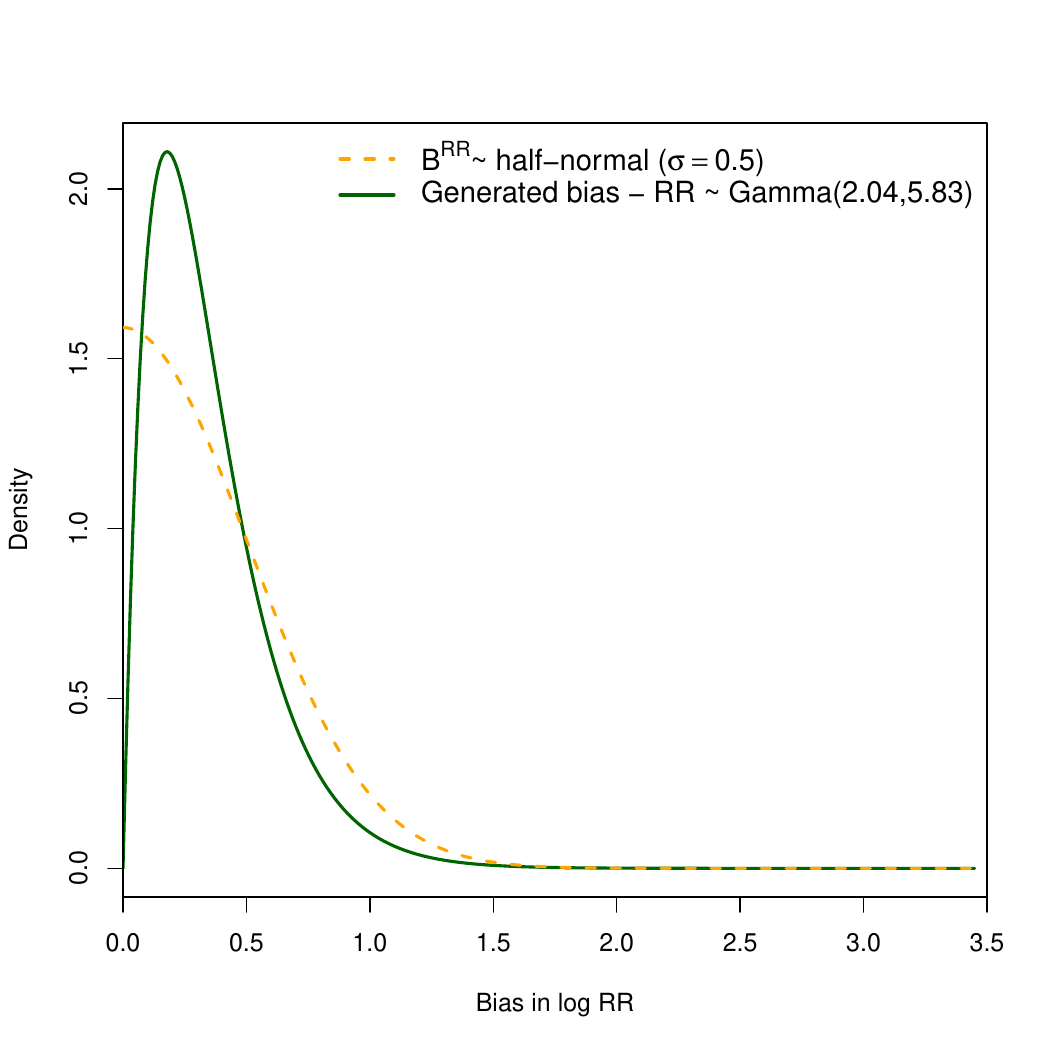}
         \caption{}
         \label{asc.prior.b}
     \end{subfigure}
     \caption{(a): Densities of prior distribution of $B_s^{OR}$  and the distribution used to generate bias terms to be added to the generated OR in biased OR studies. (b): Densities of prior distribution of $B_s^{RR}$  and the distribution used to generate bias terms to be added to the generated RR in biased RR studies.}
\end{figure}

\vspace{1in}
\begin{figure}
    \centering
    \includegraphics[width=0.6\textwidth]{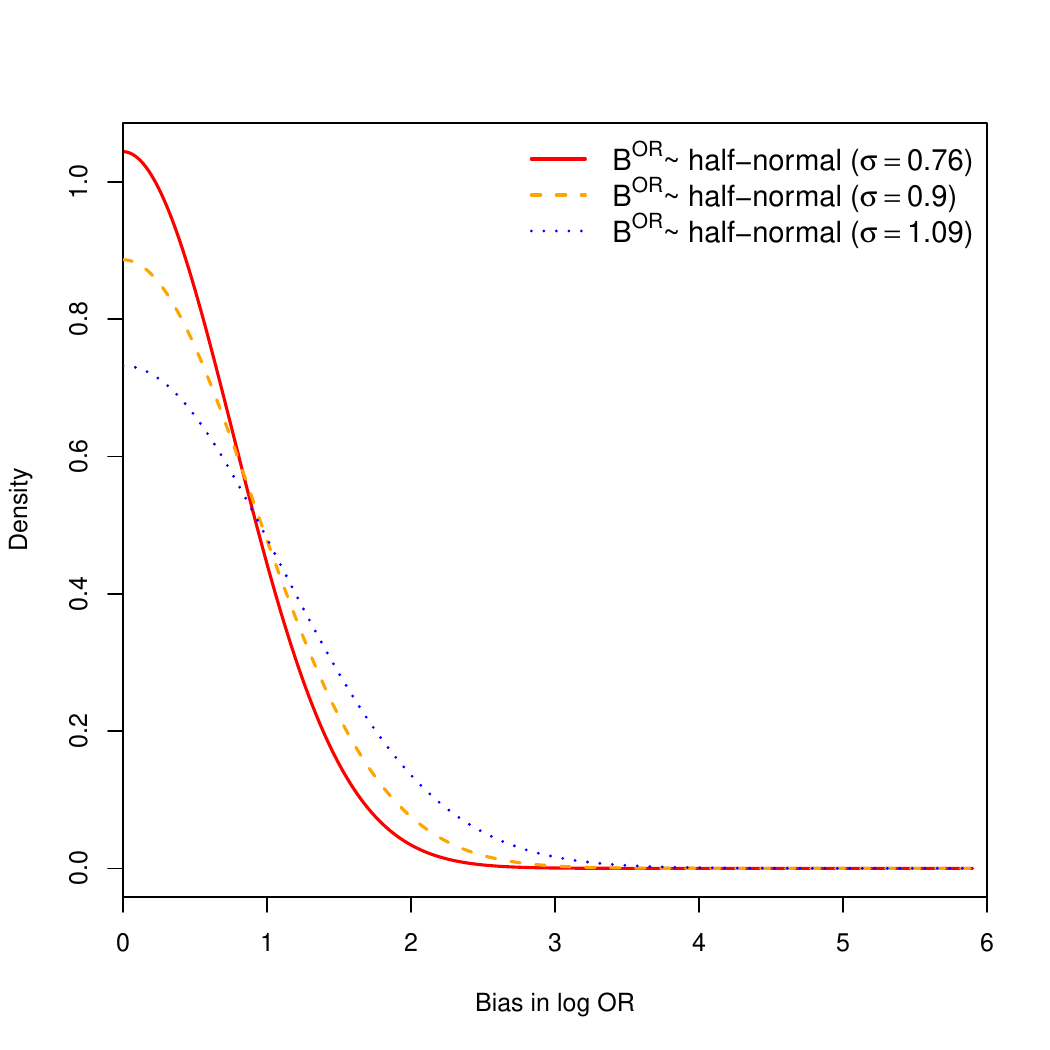}
    \caption{Densities of prior distributions of $B_s^{OR}$ used in the sensitivity analysis.}
    \label{asc.sensitivity_analysis}
\end{figure}

\begin{figure}
     \centering
    \begin{subfigure}[b]{0.45\textwidth}
        \centering
        \includegraphics[width=\textwidth]{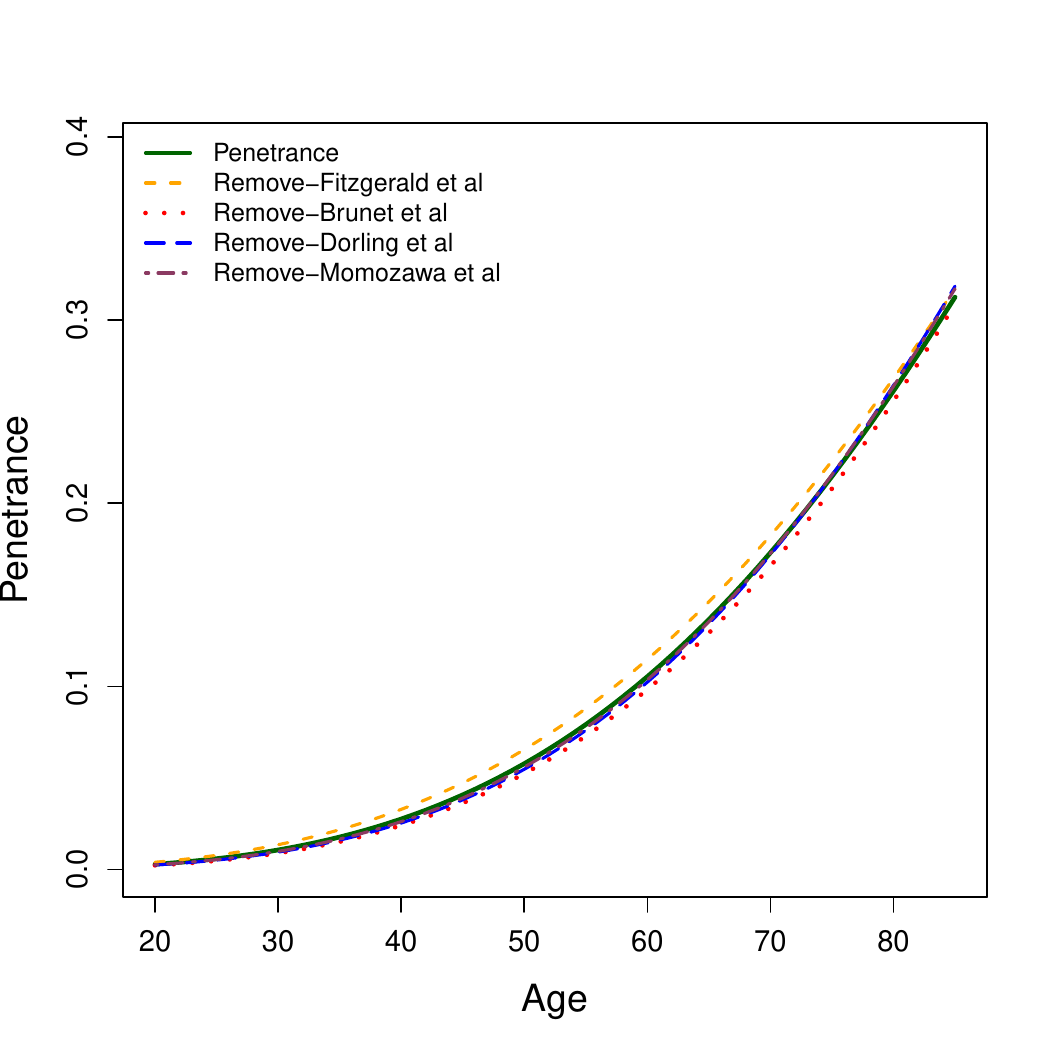}
        \caption{}
         \label{asc.sens.a}
     \end{subfigure}
     \hfill
     \begin{subfigure}[b]{0.45\textwidth}
         \centering
         \includegraphics[width=\textwidth]{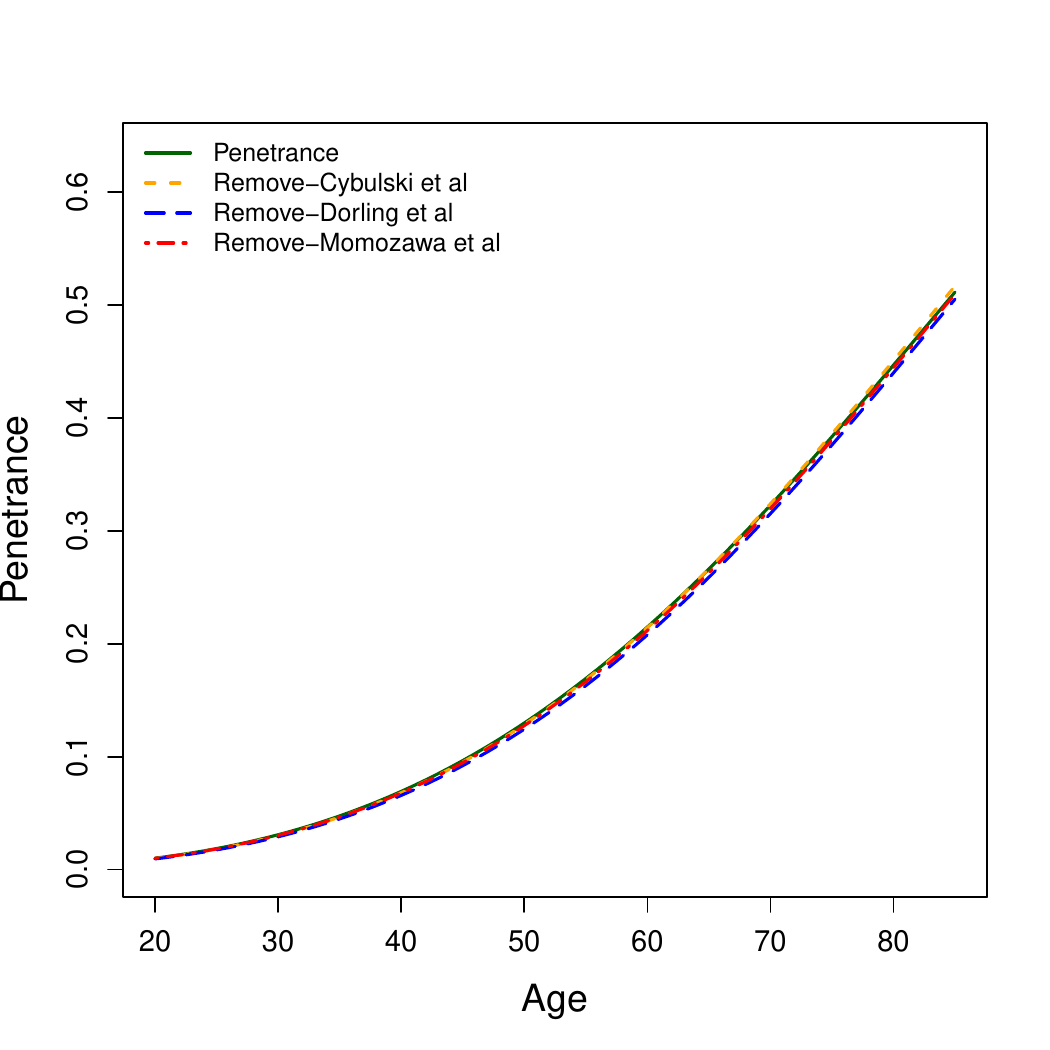}
         \caption{}
         \label{asc.sens.b}
     \end{subfigure}
     \caption{Results of sensitivity analysis: Removing studies with early onset cases or controls restricted to older ages, one at a time. (a): for ATM-BC. (b): for PALB2-BC}
         \label{asc.sens}
\end{figure}

\bibliography{supplementary}